\documentclass[usenatbib]{mn2e}
\usepackage{natbibmnfix,graphicx,times, amsmath, epsfig}

\newcommand{\uunit}{{\bf \hat{u}}}
\newcommand{\cmfast}{\textsc{\small 21CMFAST}}

\newcommand{\muKK}{\mu {\rm K^2}}
\newcommand{\PkSZ}{[\Delta^{\rm patchy}_{l3000}]^2}

\newcommand{\Ptot}{[\Delta_{l3000}]^2}

\newcommand{\xz}{({\bf x}, z)}

\newcommand{\Ts}{T_{\rm S}}
\newcommand{\Tk}{T_{\rm K}}

\newcommand{\nf}{x_{\rm HI}}
\newcommand{\avenf}{\bar{x}_{\rm HI}}

\newcommand{\lya}{Ly$\alpha$}

\newcommand{\Msun}{M_\odot}
\newcommand{\Tvir}{T_{\rm vir}}

\newcommand{\Tcmb}{T_\gamma}
\newcommand{\delT}{\delta T_b}

\newcommand{\delNL}{\delta_{\rm nl}}

\newcommand{\mfp}{R_{\rm mfp}}

\newcommand\lsim{\mathrel{\rlap{\lower4pt\hbox{\hskip1pt$\sim$}}
        \raise1pt\hbox{$<$}}}
\newcommand\gsim{\mathrel{\rlap{\lower4pt\hbox{\hskip1pt$\sim$}}
        \raise1pt\hbox{$>$}}}
\def\myputfigure#1#2#3#4#5%
{\vskip#5pt\makebox[0pt]{\hskip#2in
\includegraphics[width=#3\textwidth]{#1}}\vskip#4pt\hfill}

\newenvironment{packed_enum}{
\begin{enumerate}
  \setlength{\itemsep}{1pt}
  \setlength{\parskip}{0pt}
  \setlength{\parsep}{0pt}
}{\end{enumerate}}

\newcommand{\lowxray}{T1e4\_fuv1\_fx1\_1keV}
\newcommand{\fidxray}{T1e4\_fuv1\_fx5\_1keV}
\newcommand{\xray}{T1e4\_fuv1\_fx100\_1keV}
\newcommand{\xrayfeed}{T1e4\_fuv1\_fx100\_1keV\_feedback}
\newcommand{\matchtau}{T1e4\_fuv0.5\_fx50\_1keV}
\newcommand{\highTvir}{T1e5\_fuv1\_fx100\_1keV}
\newcommand{\highxray}{T1e5\_fuv1\_fx10000\_3keV}

\pdfoutput=1

\begin{document}

\title[X-rays in the early Universe]{Signatures of X-rays in the early Universe}

\author[Mesinger et al.]{Andrei Mesinger$^1$\thanks{email: andrei.mesinger@sns.it}, Andrea Ferrara$^1$, \& David S. Spiegel$^2$\\
$^1$Scuola Normale Superiore, Piazza dei Cavalieri 7, 56126 Pisa, Italy\\
$^2$Institute for Advanced Study, 1 Einstein Drive  Princeton, NJ, 08540, USA
}

\voffset-.6in

\maketitle

\begin{abstract}
With their long mean free paths and efficient heating of the intergalactic medium (IGM), X-rays could have a dramatic impact on the thermal and ionization history of the Universe.  Here we run several semi-numeric simulations of the Dark Ages and the Epoch of Reionization (EoR), including both X-rays and UV radiation fields, attempting to provide an intuitive framework for interpreting upcoming observations.  We explore the impact of X-rays on various signals:
(i) {\it Reionization history:} including X-rays results in an earlier, slightly more extended EoR.  However, efficient thermal feedback from X-ray heating could yield an extended epoch in which the Universe was $\approx10$\% ionized.
(ii) {\it Reionization morphology:} a sizable ($\sim$10\%) contribution of X-rays to reionization results in a more uniform morphology, though the impact is modest when compared at the same global neutral fraction, $\avenf$.  Specifically, X-rays produce a dearth of fully neutral regions and a suppression of small-scale ($k\gsim0.1$Mpc$^{-1}$) ionization power by a factor of $\lsim2$.  However, these changes in morphology cannot be countered by increasing the bias of the ionizing sources, making them a robust indicator of an X-ray contribution to the EoR.
(iii) {\it The kinetic Sunyaev-Zel'dovich (kSZ) effect:} at a fixed reionization history, X-rays decrease the kSZ power at $l=3000$ by $\approx0.5 \muKK$.  Our extreme model in which X-rays entirely drive reionization is the only one which is marginally consistent with the recent upper limits on this signal from the South Pole Telescope, assuming no thermal Sunyaev-Zel'dovich (tSZ) -- dusty galaxy cross-correlation.  Since this extreme model is unlikely, we conclude that there should be a sizable tSZ-dusty galaxy cross-correlation.
(iv) {\it The redshifted 21cm signal:} the impact of X-rays on the 21cm power spectrum during the advanced stages of reionization ($\avenf\lsim0.7$) is modest, except in extreme, X-ray dominated models.  The largest impact of X-rays is to govern the timing and duration of IGM heating.  In fact, unless thermal feedback is efficient, the epoch of X-ray heating likely overlaps with the beginning of reionization.  This results in a 21cm power spectrum which is $\sim$10--100 times higher at $\avenf\gsim0.9$ than obtained from naive estimates ignoring this overlap.  On the other hand, if thermal feedback is efficient, the resulting extended epoch between X-ray heating and reionization could provide a clean probe of the matter power spectrum in emission, at redshifts more accessible than the Dark Ages.
\end{abstract}


\section{Introduction}
\label{sec:intro}

Cosmic reionization is the last global baryonic phase change, and acts as powerful probe of early Universe physics and the first generations of galaxies.  Although the details remain unclear, it is generally accepted that reionization is mostly driven by stellar sources inside dwarf galaxies that accrete gas via molecular or atomic hydrogen line cooling. The relatively soft ultraviolet (UV) spectra of such stellar sources results in a reionization which is ``inside-out'' on large scales: ionized bubbles grow around the first,  highly clustered sources, with their eventual coalescence signaling the completion of reionization (e.g. \citealt{FZH04,McQuinn07,TC07,Zahn11}). Such UV-driven reionization scenarios result in a very inhomogeneous ionization field, where the large-scale clustering of galaxies determines the ionization morphology.

On the other hand, reionization would be noticeably different if it were driven by higher energy, X-ray photons.  Firstly, reionization morphology would be more uniform, due to the relatively large mean free path for X-rays (e.g. \citealt{McQuinn12}):
\begin{equation}
\label{eq:mfp}
\lambda_{\rm X} \approx 20 ~ \avenf^{-1} \left( \frac{E_{\rm X}}{\rm 300 eV} \right)^{2.6} \left( \frac{1+z}{10} \right)^{-2} ~ {\rm cMpc} ~,
\end{equation}
Here $\avenf$ is the mean neutral fraction of the intergalactic medium (IGM) and $E_{\rm X}$ is the photon energy.  Secondly, most of the X-ray energy gets deposited as heat when $\avenf$ surpasses a few percent (e.g. \citealt{SvS85, FS09, VEF11}), quickly raising the Jeans mass in the IGM.  Therefore, if X-rays were abundant in the early Universe, thermal feedback might delay the end stages of reionization (e.g. \citealt{RO04, KM05}).  Such a scenario could accommodate a relatively large optical depth to Thompson scattering \citep{Komatsu11} and still allow stellar UV emission to complete reionization at $z\sim6$--7, consistent with recent observations (e.g. \citealt{Bolton11, DMW11, Pentericci11, SMH12}).\footnote{However, similar extended reionization scenarios could result without X-rays: for example from strong, self-regulating feedback on small galaxies (e.g. \citealt{SH03, Ahn12, PMB12}) and/or from reionization ``stalling'' when the HII regions grow to surpass the attenuation length of ionizing UV photons \citep{FM09, Crociani11, AA12}.}

Several observations indirectly limit the allowed contribution of X-rays to reionization.  Firstly, the number density of bright active galactic nuclei (AGN) decreases sharply towards high redshifts (e.g. \citealt{MHR99, Fan01}).  Although faint AGN (so-called miniquasars) could still contribute (e.g. \citealt{Madau04, VG09}), the parameter space for such models is decreasing with more sensitive surveys probing further down the AGN luminosity function (e.g. \citealt{Willott10}).  Attention has also focused on other sources of X-rays, such as high-mass X-ray binaries (HMXBs), which dominate the X-ray contribution of star-forming galaxies (e.g. \citealt{MGS12}).  The strongest constraints on X-ray reionization comes from the present-day unresolved X-ray background (XRB).  The XRB is an indirect probe, since the observed photons originating from high-redshifts are energetic enough to have mean free paths greater than the Hubble length.  Therefore the observed photons are not the ones interacting with the IGM.  Nevertheless, estimates of the unresolved XRB from the Chandra deep fields (CDFs; \citealt{HM07}) rule out reionization models driven by hard X-ray spectra (with photon energy indices $\alpha \lsim 1$; \citealt{DHL04, McQuinn12}), provided there is no spectral break towards high-energies ($E_{\rm X} \gsim 10$ keV).

Despite the limited parameter space, there has recently been a resurgence of interest in X-ray reionization.  This has been motivated by several factors.
\vspace{-0.2cm}
\begin{packed_enum}
\item  The observed population\footnote{Given the steep faint-end slopes of the inferred luminosity functions, it is likely that galaxies below the detection thresholds contribute a significant fraction of the total ionizing background.  Therefore this motivation for X-ray reionization, although relatively popular, is not as compelling as other ones.} of high-redshift galaxies ($z\sim6$--10) seem unable to maintain an ionized Universe, baring an evolution in galaxy properties (for example, an increase with redshift in the fraction of ionizing photons capable of making it out of the galaxy; e.g. \citealt{MFC12, KF-G12, FCV12}).
\item The observed anti-correlation of HMXBs with metallicity (e.g. \citealt{Crowther10, KSG11}), as well as theoretical studies proposing a high binary fraction among the first stars \citep{TAO09, SGB10}, motivate a higher contribution from HMXBs in early galaxies \citep{Mirabel11, Fragos12}%\footnote{Note however that there is no evidence of such an evolution of the X-ray luminosity to star formation rate (SFR) ratio, up to $z\sim6$ \citep{CBH12}.  This $z\sim6$ constraint is however strongly dependent on the assumed SFR and X-ray spectral index.}.
\item The recent upper limit on the kinetic Sunyaev-Zel'dovich (kSZ) signal from the South Pole Telescope (SPT)\footnote{http://pole.uchicago.edu/} by \citet{Reichardt12} is inconsistent with stellar-driven UV reionization models \citep{MMS12}, unless there is a non-negligible correlation in the thermal SZ -- cosmic infrared background (CIB) signal.  This measurement is a probe of the ionization structure on $\sim20$ Mpc scales; X-rays with long mean free paths are a natural candidate for erasing such structure \citep{VL11, MMS12}.
\item \citet{Cappelluti12} recently detected a correlation between the CIB and the soft XRB.  The observed cross-power is best explained if the sources are at high redshifts ($z\gsim7$).
\item Recent analysis of the 4 Ms exposure of the Chandra Deep Field-South (CDF-S) by \citet{Basu-Zych12} suggests that the X-ray luminosity to star formation rate (SFR) of star forming galaxies increases approximately as $\propto(1+z)$ out to $z\sim4$.
\end{packed_enum}
\vspace{-0.2cm}

In this work, we do not attempt to predict whether or not X-rays are important to reionization (see, e.g. \citealt{DHL04, SW07, VG09, McQuinn12}).  Instead we explore various observational signatures of X-rays in the reionization and pre-ionization epochs.  We attempt to provide an intuitive framework for interpreting upcoming observations.  Given the many astrophysical uncertainties, we adopt a parametric approach, looking for general trends and physical insight.

We model X-rays in the early Universe using a public code, \cmfast.  This is a ``semi-numerical'' simulation, using well-tested and efficient algorithms to compute various cosmic fields.  Due to their long mean free paths, very large boxes (several hundred comoving Mpc) are required in order to properly model X-rays (as well as the photons responsible for pumping the 21cm line; see below).  Computing these radiation fields for the dominant galaxy population in the early Universe is computationally impractical for conventional radiative transfer (RT) algorithms.  On the other hand, analytic approaches do not preserve the spatial structure we are investigating.  Therefore semi-numerical simulations are uniquely suited for this task.

This paper is organized as follows. In \S \ref{sec:sim}, we present our simulations.  In \S \ref{sec:results} we discuss our results, including the impact of early X-rays on the evolution of the global reionization signal (\S \ref{sec:evol_reion}), reionization morphology (\S \ref{sec:morpho}), the kSZ signal (\S \ref{sec:kSZ}), and the redshifted 21cm signal (\S \ref{sec:21cm}).  Finally, in \S \ref{sec:conc}, we summarize our conclusions.

Except for fluxes, we quote all quantities in comoving units. We adopt the background cosmological parameters ($\Omega_\Lambda$, $\Omega_{\rm M}$, $\Omega_b$, $n$, $\sigma_8$, $H_0$) = (0.73, 0.27, 0.046, 0.96, 0.82, 70 km s$^{-1}$ Mpc$^{-1}$), consistent with the seven--year results of the {\it WMAP} satellite (WMAP7; \citealt{Komatsu11}).

\section{Simulations}
\label{sec:sim}

We use a parallelized version of the publicly available \cmfast\ code\footnote{http://homepage.sns.it/mesinger/Sim.html}. This code uses perturbation theory (PT) to generate the density and velocity fields, and PT plus the excursion-set formalism of \citet{FZH04} to generate the ionization fields. We summarize these calculations below.  For further details on these algorithms, see \citet{MF07}, \citet{Zahn11} and \citet{MFC11}.

\subsection{Ionization and heating by X-ray photons}

In the early stages of reionization, there is a small HII region local to each source (e.g. \citealt{Cen06}), whereas the X-rays are able to penetrate deep into the mostly-neutral IGM.  Outside of the local HII regions, we compute the temperature, $T_{\rm K}$, and ionized fraction, $x_e$, at a given redshift, $z$, and position in our simulation box, ${\bf x}$, according to:

\begin{equation}
\label{eq:ion_rateacc}
\frac{dx_e\xz}{dz} = \frac{dt}{dz} \left[ \Gamma_{\rm ion}
  - \alpha_{\rm A} C x_e^2 n_b f_{\rm H} \right] ~ ,
\end{equation}

\begin{align}
\label{eq:dTkdzacc}
\nonumber \frac{d\Tk\xz}{dz} &= \frac{2}{3 k_B (1+x_e)} \frac{dt}{dz} \sum_p \epsilon_p \\
&+ \frac{2 \Tk}{3 n_b} \frac{dn_b}{dz}  - \frac{\Tk}{1+x_e} \frac{dx_e}{dz} ~ ,
\end{align}

\noindent where $n_b=\bar{n}_{b, 0} (1+z)^3 [1+\delNL\xz]$ is the total (H + He) baryonic number density at $\xz$, $\epsilon_p\xz$ is the heating rate per baryon for process $p$ in erg s$^{-1}$, $\Gamma_{\rm ion}$ is the ionization rate per baryon, $\alpha_{\rm A}$ is the case-A recombination coefficient, $C\equiv \langle n^2 \rangle / \langle n \rangle^2$ is the clumping factor on the scale of the simulation cell (here taken to be $C=2$), $k_B$ is the Boltzmann constant, $f_{\rm H}$ is the hydrogen number fraction. 

For X-ray heating and ionization, we have:

\begin{equation}
\label{eq:eps_explicit}
\epsilon_{\rm X}\xz = \int_{\rm Max[\nu_0, \nu_{\tau=1}]}^\infty d\nu \frac{4\pi J}{h \nu} \sum_i (h\nu - E^{\rm th}_i)  f_{\rm heat} f_i x_{i}  \sigma_i
\end{equation}

and

\begin{equation}
\label{eq:lam}
\Gamma_{\rm ion}\xz = \int_{\rm Max[\nu_0, \nu_{\tau=1}]}^\infty d\nu \frac{4\pi J}{h \nu} \sum_i f_i x_{i}  \sigma_i F_i ~ ,
\end{equation}
\begin{align}
\nonumber F_i = \left(h\nu - E^{\rm th}_i \right) \left( 
\frac{f_{\rm ion, HI}}{ E^{\rm th}_{\rm HI}} + 
\frac{f_{\rm ion, HeI}}{ E^{\rm th}_{\rm HeI}} + 
\frac{f_{\rm ion, HeII}}{E^{\rm th}_{\rm HeII}} \right)
 + 1
\end{align}

\noindent where the frequency integral includes the contribution from photons within a mean free path of $({\bf x}, z)$\footnote{Note that this spatially-averaged, frequency-dependent opacity is computed including the total ionized fraction from both X-rays and UV photons \citep{MFC11}. Our formalism assumes that UV photons, with their short mean free paths, carve out fully-ionized HII regions in the IGM (e.g. \citealt{Zahn11}).  If we denote the covering factor of these HII regions as $Q_{\rm HII}$, then the total ionized fraction (including the partial ionizations from X-rays in eq. \ref{eq:ion_rateacc}), can be written as $\bar{x}_i \approx Q_{\rm HII} + (1-Q_{\rm HII}) x_e$.  Similarly, we denote the mean neutral fraction as $\avenf = 1-\bar{x}_i$. The X-ray optical depth between $z$ and $z'$ can then be written as $\tau_{\rm X}(\nu, z, z') = \int_{z'}^{z} d\hat{z} (cdt/d\hat{z}) (1-Q_{\rm HII}) \bar{n}_b \tilde{\sigma}$, where the photo-ionization cross-section is weighted over species,$\tilde{\sigma}(z, \hat{\nu}) \equiv f_{\rm H} (1-\bar{x}_e) \sigma_{\rm H} + f_{\rm He} (1-\bar{x}_e) \sigma_{\rm HeI} + f_{\rm He} \bar{x}_e \sigma_{\rm HeII}$ and is evaluated at $\hat{\nu} = \nu (1+\hat{z})/(1+z)$.}, and we perform a 
 sum over species, $i=$ HI, HeI, or HeII, in which $f_i$ is the number fraction, $x_i$ is the cell's species ionization fraction [which for HI and HeI is $(1-x_e)$, and for HeII is $x_e$], $\sigma_i$ the ionization cross-section, and $E^{\rm th}_i$ is the ionization threshold energy of species $i$. Furthermore, $f_{\rm heat}[h\nu - E^{\rm th}_i, x_e({\bf x, z})]$ and $f_{\rm ion, j}[h\nu - E^{\rm th}_i, x_e({\bf x, z}), j]$ are the fraction of the primary electron's energy going into heat and secondary ionizations of species $j$ respectively (taken from \citealt{FS09}).

In the above equations, the angle-averaged specific intensity, $J(\nu, {\bf x}, z)$, (in erg s$^{-1}$ Hz$^{-1}$ cm$^{-2}$ sr$^{-1}$) can be computed integrating along the light-cone:

\begin{equation}
\label{eq:J}
J(\nu, {\bf x}, z) = \frac{(1+z)^3}{4\pi} \int_{z}^{\infty} dz' \frac{c dt}{dz'} \epsilon_{h \nu} ~ ,
\end{equation}

\noindent and the comoving specific emissivity is evaluated at $\nu_e = \nu (1+z')/(1+z)$:

\begin{align}
\label{eq:emissivity}
\nonumber \epsilon_{h \nu}(\nu_e, {\bf x}, z') = &\alpha h \frac{N_{\rm X}}{\mu m_p} \left( \frac{\nu_e}{\nu_0} \right)^{-\alpha} \\
&\left[ \rho_{\rm crit, 0} \Omega_b f_\ast (1+\bar{\delta}_{\rm nl}) \frac{d f_{\rm coll}}{dt} \right] ~ ,
\end{align}
\noindent where $N_{\rm X}$ is the number of X-ray photons per stellar baryon, $\mu m_p$ is the mean baryon mass, $\rho_{\rm crit, 0}$ is the current critical density, $f_\ast$ is fraction of baryons converted into stars (we take $f_\ast=0.1$), $\bar{\delta}_{\rm nl}$ is the mean non-linear overdensity and $f_{\rm coll}$ is the fraction of matter collapsed into halos with virial temperatures greater than $\Tvir$, which is computed according to the hybrid prescription of \citet{BL08}. The quantity in the brackets is the comoving star formation rate density (SFRD).  The specific emissivity is chosen to have a spectral (energy) index of $\alpha=1.5$ both in order to facilitate comparisons with previous works (e.g. \citealt{PF07, Santos08, Baek10, MFC11}), and also since a significant contribution from harder spectra is ruled out by observations of the XRB \citep{McQuinn12} unless there is a break at high energies ($E_{\rm X}\gsim10$ keV).

\subsection{Ionization by UV photons}
\label{sec:UV}

The contribution of UV photons to reionization is computed as a post-processing step, following the X-rays. We use the excursion-set prescription of \citet{FZH04}, slightly modified according to \citet{Zahn11} (our 'FFRT' scheme).  This algorithm determines whether a region is ionized by comparing the (integrated) number of ionizing photons to the number of neutral atoms.
Specifically, a simulation cell at coordinate ${\bf x}$ is flagged as ionized if 
\begin{equation}
\label{eq:HII_barrier}
\zeta_{\rm UV} f_{\rm coll}({\bf x}, z, R) \geq 1-x_e({\bf x}, z, R) ~ ,
\end{equation}
\noindent where $\zeta_{\rm UV}$ is an ionizing efficiency parameter (described below), and as above, $f_{\rm coll}$ is the fraction of mass residing in dark matter halos inside a sphere of radius $R$ and mass $M=4/3 \pi R^3 \rho$, where $\rho = \bar{\rho} [1+\bar{\delta}_{\rm nl}]$.  In this work we include on the RHS the mean fraction of neutral hydrogen remaining within $R$: $1-x_e$, computed according to eq. (\ref{eq:ion_rateacc}). 

As for the X-rays, this tabulation includes all halos above a set virial temperature threshold, $\Tvir$.  We do not explicitly compute individual halo locations (as in \citealt{MF07}), but instead apply the conditional excursion-set formalism directly on the non-linear density field to compute $f_{\rm coll}$ \citep{Zahn11}.   The ionization field is also computed following the excursion-set approach (e.g., \citealt{Bond91, LC93, FZH04}). We decrease the filter scale $R$, starting from some maximum value $R_{\rm max}$.  A physical choice for $R_{\rm max}$ corresponds to the mean free path of ionizing photons, $\mfp$ \citep{FO05, CHR09, AA12}. We take $R_{\rm max}=40$ Mpc, consistent with recent observations (e.g. \citealt{SC10}) and theoretical estimates (e.g. \citealt{MOF11}).
 If at any $R$ the criterion in equation (\ref{eq:HII_barrier}) is met, this cell is flagged as ionized.

%Note that this formalism assumes that UV photons, with their short mean free paths, carve out fully-ionized HII regions in the IGM (e.g. \citealt{Zahn11}).  If we denote the covering factor of these HII regions as $Q_{\rm HII}$, then the total ionized fraction (including the partial ionizations from X-rays in eq. \ref{eq:ion_rateacc}), can be written as $\bar{x}_i \approx Q_{\rm HII} + (1-Q_{\rm HII}) x_e$.  Similarly, we denote the mean neutral fraction as $\avenf = 1-\bar{x}_i$.

\subsection{The runs}
\label{sec:runs}

Our simulation boxes are 750 Mpc on a side, with a resolution of 500$^3$.  We integrate the evolution from $z=40$ down to $z=5.6$, including the end of the Dark Ages and reionization.  Here we discuss the astrophysical parameters we vary, and motivate some fiducial choices.

We parameterize the ionizing efficiency of UV photons as follows:

\begin{equation}
\label{eq:f_UV}
f_{\rm UV} \equiv  \frac{\zeta_{\rm UV}}{30} \approx \bigg(\frac{N_\gamma}{4400}\bigg) \bigg(\frac{f_{\rm esc}}{0.1}\bigg) \bigg(\frac{f_\ast}{0.1}\bigg) \bigg(\frac{1.5}{1+\bar{n}_{\rm rec}}\bigg) ~ ,
\end{equation}

\noindent where $f_{\rm esc}$ is the fraction of UV ionizing photons that escape into the IGM, $N_\gamma$ is the number of ionizing photons per stellar baryon, and $\bar{n}_{\rm rec}$ is the mean number of recombinations per baryon (a factor which appears since $\zeta_{\rm UV}$ is defined after integrating the ionization rate over cosmic time).  A value of $f_{\rm UV}\approx1$ agrees with the observed value of $\tau_e\approx0.09$ \citep{Komatsu11}, in the absence of an X-ray contribution to reionization.  Although our models depend only on the product in eq. (\ref{eq:f_UV}), we show on the RHS some reasonable values for the component terms.  $N_\gamma=4400$ is expected for PopII stars (e.g. \citealt{BL05_WF}), and studies of the high-redshift Lyman alpha forest suggest $\bar{n}_{\rm rec} \sim 0$ \citep{Miralda-Escude03, BH07, MOF11}.  On the other hand, the parameters $f_{\rm esc}$ and  $f_\ast$ are extremely uncertain in high-redshift galaxies  (e.g. \citealt{GKC08, WC09, FL12}), though their product might be more robust \citep{WC09, Paardekooper11}.

Similarly, the X-ray efficiency is parameterized as follows:

\begin{equation}
\label{eq:f_X}
f_{\rm X} \equiv \bigg(\frac{N_{\rm X}}{0.2}\bigg) \bigg(\frac{f_\ast}{0.1}\bigg) ~ .
\end{equation}

\noindent The value of $N_{\rm X}=0.2$ X-ray photons per stellar baryon corresponds to a total X-ray luminosity above $h\nu_0=0.3$ keV of $L_{\rm X, {\rm 0.3+keV}}\approx10^{40}$ erg s$^{-1}$ ($\Msun$ yr$^{-1})^{-1}$, using our spectral energy index of $\alpha=1.5$.  This choice is consistent with (a factor of $\approx$2 higher than) an extrapolation from the 0.5--8 keV measurement of \citet{MGS12}, $L_{\rm X, {\rm 0.5-8keV}}\approx3\times10^{39}$ erg s$^{-1}$ ($\Msun$ yr$^{-1})^{-1}$.  \citet{MGS12} find a factor of 2--3 intrinsic scatter in this relation for their sample of star-forming galaxies\footnote{Our fiducial choice of X-ray emissivity is motivated by local observations of star-forming galaxies which dominate the unresolved X-ray background.  However, our results are also valid for other sources of X-rays, such as mini quasars, with the corresponding emissivity.}, while other studies in different bands have estimates which are also within a factor of few of this value (e.g. \citealt{GGS04, RCS03, Persic04, PR07, Lehmer10})\footnote{We note that our fiducial choice ($f_{\rm X}=1$) for the X-ray efficiency, $N_{\rm X}$, is a factor of $\approx4$ lower than that used in prior studies, based on an extrapolation of the earlier 2--10keV measurement by \citet{GGS04} down to $h\nu_0=0.2$ keV.  Therefore our $f_{\rm X}=5$ model below is a closer match to most prior theoretical studies (e.g. \citealt{Furlanetto06, PF07, Santos08, WGW09, MW11})}.  
Recently, the 4 Ms exposure of the Chandra Deep Field-South (CDF-S; e.g. \citealt{Xue11}) has allowed us to study the redshift evolution of this relation.  There is some evidence of evolution out to $z\sim4$ (\citealt{Basu-Zych12}; though see \citealt{CBH12}).  However, interpreting the derived upper limits on the X-ray flux of $z\sim6$ galaxies is very uncertain, and in particular is heavily dependent on the assumed extrapolation to the soft X-rays which are the ones actually interacting with the IGM.  For example, the 2$\sigma$ upper limits from \citet{CBH12} still allow $N_{\rm X}\sim1000$ for their $z\sim6$ galaxy sample, e.g. assuming a SFR of 0.1 $\Msun$ yr$^{-1}$, $h\nu_0=0.3$ keV and $\alpha=1.5$.
%Recently, using the 4 Ms exposure of the Chandra Deep Field-South (CDF-S) field, \citet{CBH12} set upper limits on the rest-frame $\sim$3.75--15 keV flux of $z\sim6.5$ galaxies, finding no evidence of redshift evolution (see also \citealt{Xue11}).  We note however that their 2$\sigma$ upper limits still allow $N_{\rm X}\sim1000$ for their CDF-S $z\sim6$ LBG sample, e.g. assuming a SFR of 0.1 $\Msun$ yr$^{-1}$, $h\nu_0=0.3$ keV and $\alpha=1.5$.

The strongest constraint on the X-ray contribution to reionization comes from the unresolved soft XRB observed in the CDFs.  As mentioned in the introduction, the XRB is an indirect probe since the observed photons are not the ones interacting with the IGM.  Therefore, constraints are sensitive to the assumed spectral index or any spectral break at high energies.  \citet{HM07} estimate the 1--2 keV unresolved XRB to be 3.4 $\pm$ 1.4 $\times$ 10$^{-13}$ erg s$^{-1}$ cm$^{-2}$ deg$^{-2}$.  This can be compared with the XRB at $z=0$ in our models, estimated from eq. (\ref{eq:J}).  X-rays in our fiducial model below make a negligible contribution to the XRB, contributing 0.4\% (5\%) of the unresolved XRB if the sources remain on until $z\approx$ 10 (6).
%1.2e-15  1.7e-14
On the other hand, in our $f_{\rm X}=10^4$ ``extreme'' model where X-rays dominate reionization (see below), the sources at $z\gsim10$ already saturate the unresolved XRB.  This is roughly consistent with recent simple estimates (c.f. Fig. 1 in \citealt{McQuinn12}; see also \citealt{DHL04, SHV07}).

Following previous works (e.g. \citealt{Furlanetto06, PF07, Santos08, WGW09, Baek10, MFC11}), we assume a simple obscuration model for our X-ray sources, with a step function at $h \nu_0$.  This allows us to separately treat the sources of X-ray and UV photons, isolating the imprint of the former.  On the other hand, the stellar-driven UV is assumed to be soft enough that each photon above the Lyman limit produces a single ionization (eq. \ref{eq:HII_barrier}).
Although sources of X-rays in principle do also contribute some UV photons, local observations find that the majority of both AGN and HMXBs are significantly obscured (e.g. \citealt{Lutovinov05, Tozzi06}).  Photons below our fiducial choice of $h \nu_0=$ 300 eV have optical depths exceeding unity for $N_{\rm HI}\gsim10^{21.5}$ cm$^{-2}$, consistent with the column densities seen in high-redshift gamma-ray bursts (GRBs) \citep{Totani06, Greiner09}.

The final free parameter we vary is $\Tvir$, the minimum virial temperature of halos hosting galaxies.  Although our formalism allows for disparate values of $\Tvir$ for X-ray and UV sources, for simplicity we use the same value for both populations. Our fiducial choice of $\Tvir=10^4$ K corresponds to the atomic cooling threshold. Although the first stars are likely hosted by smaller halos with $\Tvir\sim$few$\times10^3$ K (e.g. \citealt{HTL96, ABN02, BCL02}), star formation inside such small halos was likely inefficient (with a handful of stars per halo), and was likely suppressed by heating and feedback processes (\citealt{HAR00,RGS01,MBH06, HB06, TH10, OM12}; though \citealt{Ahn12} suggest the feedback could be self-regulating).  We therefore do not include radiation from such so-called minihalos.
  $\Tvir$ could also have been larger than $10^4$K:  radiative and/or mechanical feedback (e.g. \citealt{SH03}) eventually suppresses star formation inside $\Tvir \lsim 10^5$ K halos, though the details and timing of these processes are not well understood at high redshifts (e.g. \citealt{MD08, OGT08, PS09})\footnote{ It is unlikely that $\Tvir$ is much greater than few$\times10^5$~K, since these values approximately correspond to the faint end of the observed galaxy luminosity functions at $z\sim6$ (e.g. \citealt{Bouwens08, Labbe10, SFD11, FDO11}).  Additionally, high $\Tvir$ models have difficulty in latching onto the slow evolution of the observed emissivity, as inferred from the \lya\ forest at $3<z<6$ (see Fig. 12 in \citealt{MMS12}).}.

In summary, the total (integrated) number of photons per stellar baryon escaping into the IGM in our models is given by $N_\gamma f_{\rm esc}$ for the ionizing UV photons, and $N_{\rm X}$ for the X-rays.  For the UV we don't specifically adopt a spectral energy distribution (SED), instead assuming that the spectra are soft enough that the majority of photons will ionize at most one atom.  For the X-rays, we assume a rest-frame luminosity, $L_e \propto \nu^{1.5}$, extending from $\nu=\nu_0$ to $\nu=\infty$.

%%%%%%%%%%%%%%%%%%%%% Parameter table %%%%%%%%%%%%%%%%
\begin{table*}
\begin{tabular}{|c|c|c|c|c|c|c|c|}
\hline
Run name & $T_{\rm vir}$ ($10^4$ K) & $f_{\rm UV}$ & $f_{\rm X}$ & $h\nu_0$ (keV) & $\tau_e$ & X-ray fraction & $\PkSZ$ ($\muKK$)\\
\hline
\lowxray  & 1  & 1    & 1 & 0.3 &  0.092 %0.088
 & 3\% %0.02
 & 2.3  \\
\fidxray  & 1  & 1    & 5 & 0.3 &  0.094 %0.089
 & 8\% &  2.3  \\
\xray     & 1  & 1    & 100 & 0.3 & 0.107 %0.103
 & 46\% %0.48
 & 2.1  \\
\xrayfeed & $1\rightarrow10$  & 1    & 100 & 0.3 & 0.069 %0.066
 & 46\% %0.48
 & 1.3  \\
\matchtau & 1  & 0.5 & 50 & 0.3 & 0.092 %0.088
 & 45\% %0.47
 & 1.8  \\
\highTvir & 10 & 1    & 100    & 0.3 & 0.063 %0.060
 & 48\% %0.49
 &  1.2  \\
\highxray & 10 & 1    & 10$^4$  & 0.9 & 0.108 %0.108
 & 96\% %0.98
 &  0.95  \\
\hline
\end{tabular}
\caption{Model parameters and related observational signatures, as described in the text.  %The X-ray fraction is an estimate of the fraction of total ionizations contributed by X-rays in each model. 
%$^\ast$ The "X-ray fraction" is an estimate of the fractional contribution of X-rays to reionization.  We estimate this quantity by comparing what would be the mean ionized fraction in the absence of UV photons, $\bar{x}_{i, X}$, to the actual ionized fraction, $\bar{x}_i$, when the universe is half ionized.  In other words the X-ray fraction is $2 \bar{x}_{i, X}(\bar{x}_i=0.5)$.  We find that this estimate is not particularly sensitive to the epoch when this comparison is made, since X-ray reionization occurs slowly compared to UV reionization.
\label{tbl:runs}}
\end{table*}
%%%%%%%%%%%%%%%%%%%%%%%%%%%%%%%%%%%%%%%%%%%%%%%%%%%%%%

%%%%%%%%%%%%%%%%%%%%%%%%%%%%%%%%%%%%%%%%%%%%%%%%%%%%%%%%%%%%%%%%%%%%%
\begin{figure*}
\vspace{+0\baselineskip}
{
\includegraphics[width=0.9\textwidth]{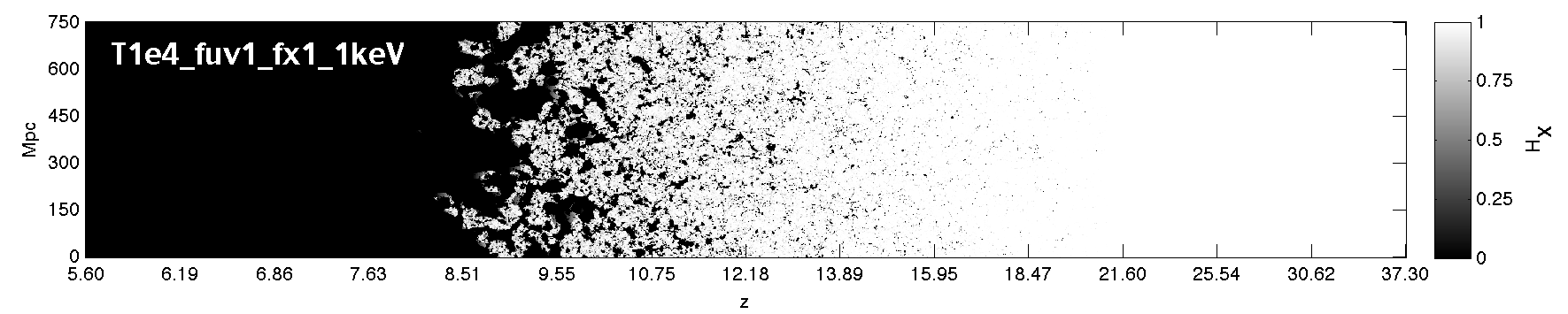}
\includegraphics[width=0.9\textwidth]{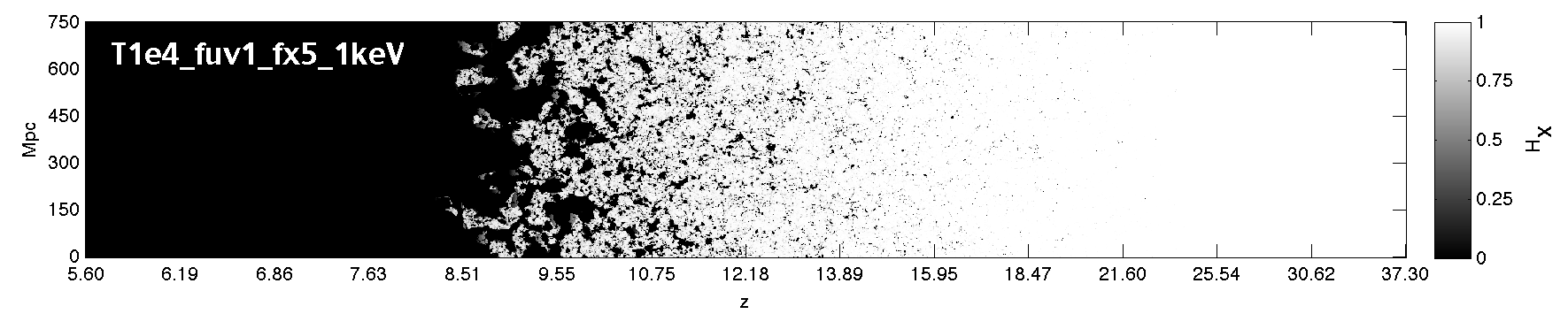}
\includegraphics[width=0.9\textwidth]{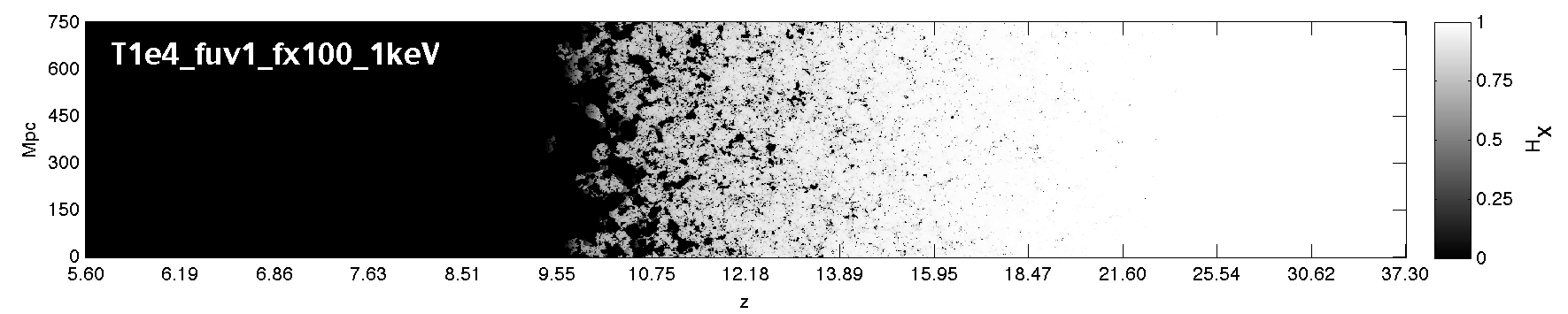}
\includegraphics[width=0.9\textwidth]{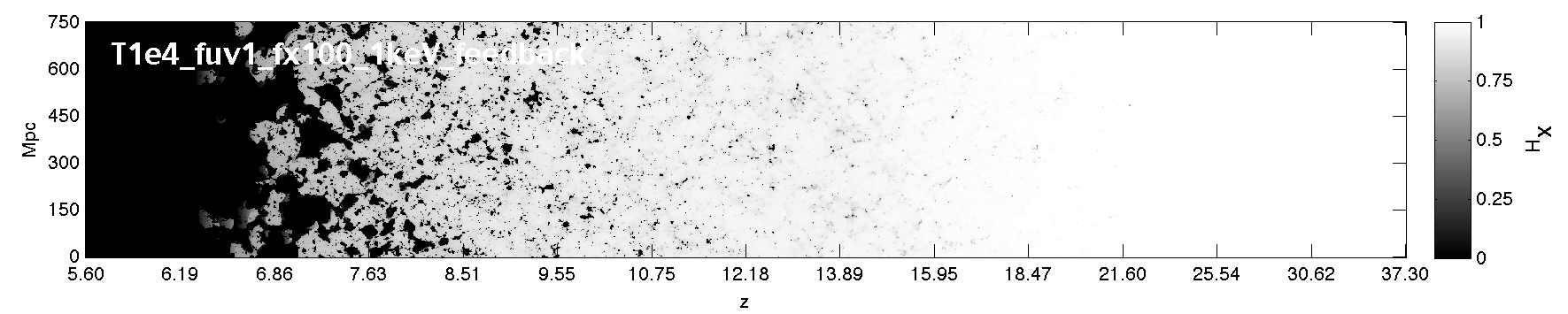}
\includegraphics[width=0.9\textwidth]{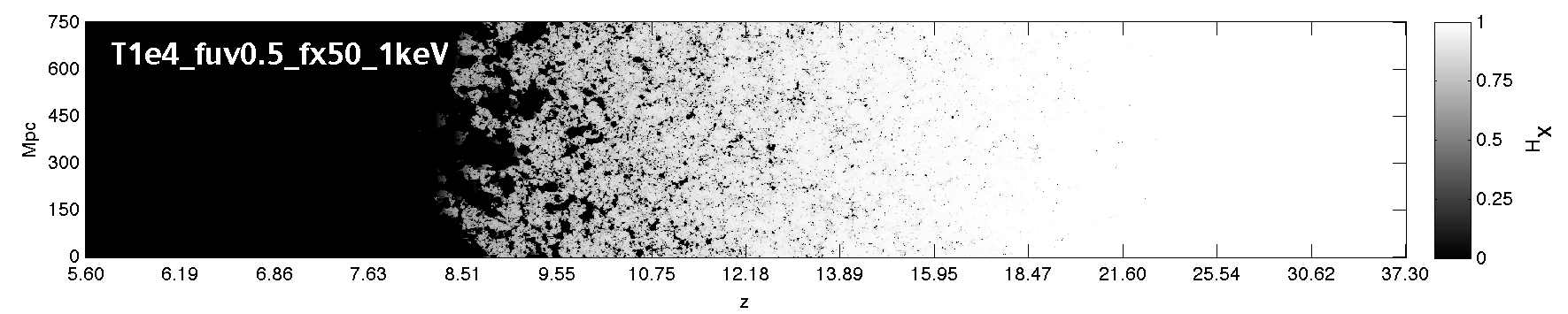}
\includegraphics[width=0.9\textwidth]{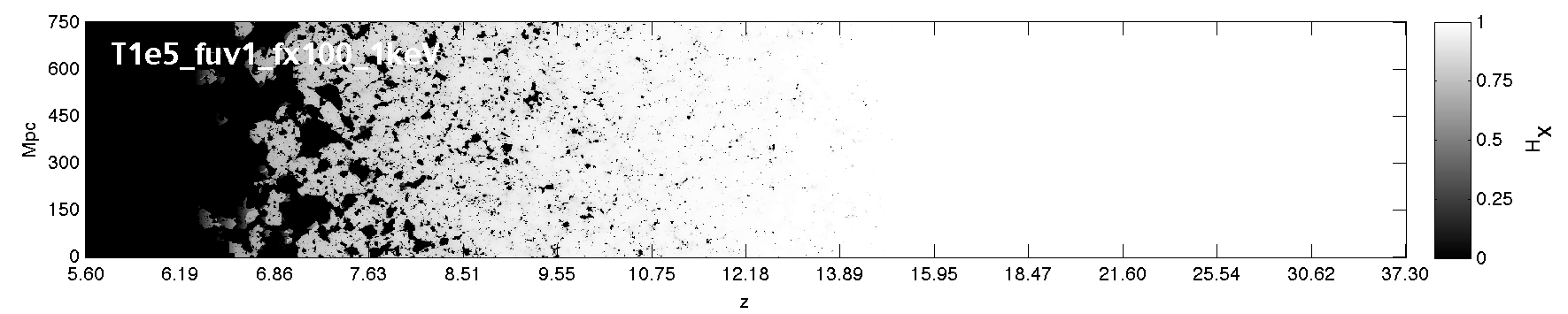}
\includegraphics[width=0.9\textwidth]{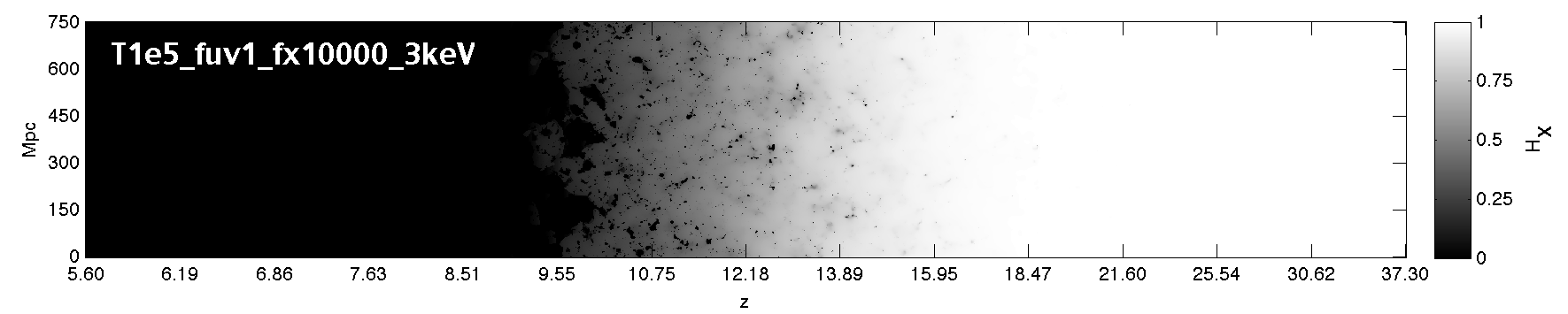}
}
\caption{
Slices through ionization fields, showing evolution with comoving distance along the light cone (x-axis). All slices are 1.5 Mpc (1 cell) deep.
\label{fig:xH_lightcone}
}
\vspace{-1\baselineskip}
\end{figure*}
%%%%%%%%%%%%%%%%%%%%%%%%%%%%%%%%%%%%%%%%%%%%%%%%%%%%%%%%%%%%%%%%%%%%%

We list all of our runs below (summarized also in Table \ref{tbl:runs}):

\begin{itemize}
\item {\bf \lowxray} -- This is our ``fiducial'' run, with $\Tvir=10^4$ K, $f_{\rm UV}=1$, $f_{\rm X}=1$, and $h\nu_0$= 0.3 keV, resulting in a spectrally-averaged mean photon energy $h\bar{\nu}=$ 0.9 keV.  This model predicts $\tau_e=0.092$\footnote{Interestingly, the fact that reionization is ``inside-out'' on large scales results in a different $\tau_e$ then would be estimated assuming homogeneous reionization.  This is due to the fact that $\tau_e$ is proportional to the sightline-averaged integral of the ionized fraction times the density: $\langle x_{i}\times n \rangle$, and $\langle x_{i} \times n \rangle$ $\neq$ $\langle x_{i} \rangle \times \langle n \rangle$.  In patchy reionization models, where the high-redshift overdensities hosting early galaxies ionize first, this inequality results in values of $\tau_e$ that are a few percent higher than estimated assuming $x_{i}$ and $n$ are uncorrelated (e.g. $\tau_e=0.092$ instead of $\tau_e=0.088$ in our fiducial model).  Understandably, this bias decreases as reionization becomes more homogeneous, such that in our ``extreme'' model $\tau_e$ has the same value, 0.11, computed both ways.  The correlation of the ionization and density fields can have a $\lsim10$\% effect on other global signals, such as the 21cm brightness temperature during the advanced stages of reionization (Wheeler et al., in preparation).  We therefore stress that the correlations between various cosmic fields should be taken into account for  precise estimates of even globally-averaged signals.}, consistent with WMAP7.
In this fiducial case, UV photons dominate reionization, with X-rays accounting for only $\approx3$\% of the total ionizations.%\footnote{Our parameter choices fix the X-ray and UV emissivity; however separating out their ionizing contributions is not straightforward.  We estimate the X-ray contribution to reionization by taking the value of $\bar{x}_e$ after the completion of reionization.  Since the UV ionization contribution is done by post-processing the X-ray generated $x_e$ field, this estimate is accurate in the limit of uniform X-ray ionizations (i.e. a homogeneous $x_e$ field).

\item {\bf \fidxray} -- This run differs from the fiducial only in having an X-ray efficiency which is a factor of $\approx$5 higher, making it more in line with several previous works (e.g. \citealt{Furlanetto06, PF07, Santos08, WGW09, MW11, Valdes12}).

\item {\bf \xray} -- Here we increase the X-ray efficiency by a factor of 100 from our fiducial choice.  In this model, X-rays ionize approximately half of the IGM\footnote{The relative contribution of UV and X-ray photons to reionization can also be estimated from their relative number of ionizations per stellar baryon: $N_\gamma  f_{\rm esc}$ for the UV, and $N_{\rm X}  N_i$ for the X-rays.  Here $N_i$ is the mean number of ionizations per X-ray photon.  Depending on the spectrum and ionized fraction, secondary ionizations can drive $N_i\approx$10--30 early in reionization, transitioning to $N_i=1$ late in reionization when all of the primary electron's energy gets deposited as heat.  Using $N_i\approx20$ and the parameter values in our \xray\ model, we confirm that the number of ionizations by X-rays and UV photons are comparable: $N_{\rm X}  N_i \sim 400 \sim N_\gamma  f_{\rm esc}$.}.  We use this model as a reference scenario in which X-rays are important, and vary additional parameters below.

\item {\bf \xrayfeed} -- This model is the same as \xray\, but includes a simple, extreme prescription for thermal feedback, with $\Tvir$ increasing to 10$^5$ K, when the mean temperature of the IGM surpasses $10^4$ K. Although thermal (i.e. radiative feedback) was initially suspected of being important (e.g. \citealt{SGB94, TW96, Gnedin00filter}), later studies concluded that its impact was smaller at high-redshifts (e.g. \citealt{RGS02b, KM05, MD08, OGT08}).  Since the \xrayfeed\ model is extreme in that it includes an {\it instantaneous} transition to a {\it high} $\Tvir$ value, it serves to bracket the expected effects of feedback.

\item {\bf \matchtau} -- X-rays also contribute about 1/2 of the ionizations in this model, but the efficiencies of both the UV and X-rays have been lowered by a factor of $\approx0.52$.  This is done in order to delay reionization so that $\tau_e$ is the same as in our fiducial run, facilitating comparison.

\item {\bf \highTvir} -- This model has $f_{\rm X}=100$ as in \xray, but with $\Tvir=10^5$ K throughout cosmic time, corresponding to inefficient star formation in smaller dwarf galaxies.

\item {\bf \highxray} -- This is our ``extreme'' model, in which high energy X-rays drive reionization (accounting for over 96\% of all ionizations).  Specifically, we take $f_{\rm X}=10^4$, $\Tvir=10^5$ K, and $h\nu_0=0.9$ keV (requiring $N_{\rm HI}\gsim10^{23}$ cm$^{-2}$, and resulting in a spectrally-averaged photon energy of $h\bar{\nu}=2.7$ keV).  As mentioned above, the unresolved XRB is saturated by $z\gsim10$ in this model.
\end{itemize}

%Before presenting our results, we estimate the anticipated trends by looking at the mean free path of X-ray photons...
%mfp weighted by the spectrum as a function of $h\nu_0$
%do at xH=1 and xH=0.7
%do at z=10, 20
%do for alpha=-1.5 =-0.5
%note additional weighting by the light cone: i.e. structure grows exponentially.

\section{Results}
\label{sec:results}

\subsection{Evolution and Timing of Reionization}
\label{sec:evol_reion}

In Fig. \ref{fig:xH_lightcone}, we show slices through ionization fields, showing evolution with comoving line-of-sight (LOS) distance along the light cone (x-axis).  X-rays have a negligible contribution to reionization in the fiducial model.  The morphology remains unaffected as the X-ray efficiency, $f_{\rm X}$, is increased by a factor of five.  However, as the X-ray efficiency is increased further, one starts to notice several differences in timing and morphology.  As expected, reionization occurs earlier, has less small-scale structure, with a partially-ionized homogeneous ``haze'' developing in the early stages of reionization.  We quantify these trends below.

\begin{figure}
\vspace{-1\baselineskip}
{
\includegraphics[width=0.45\textwidth]{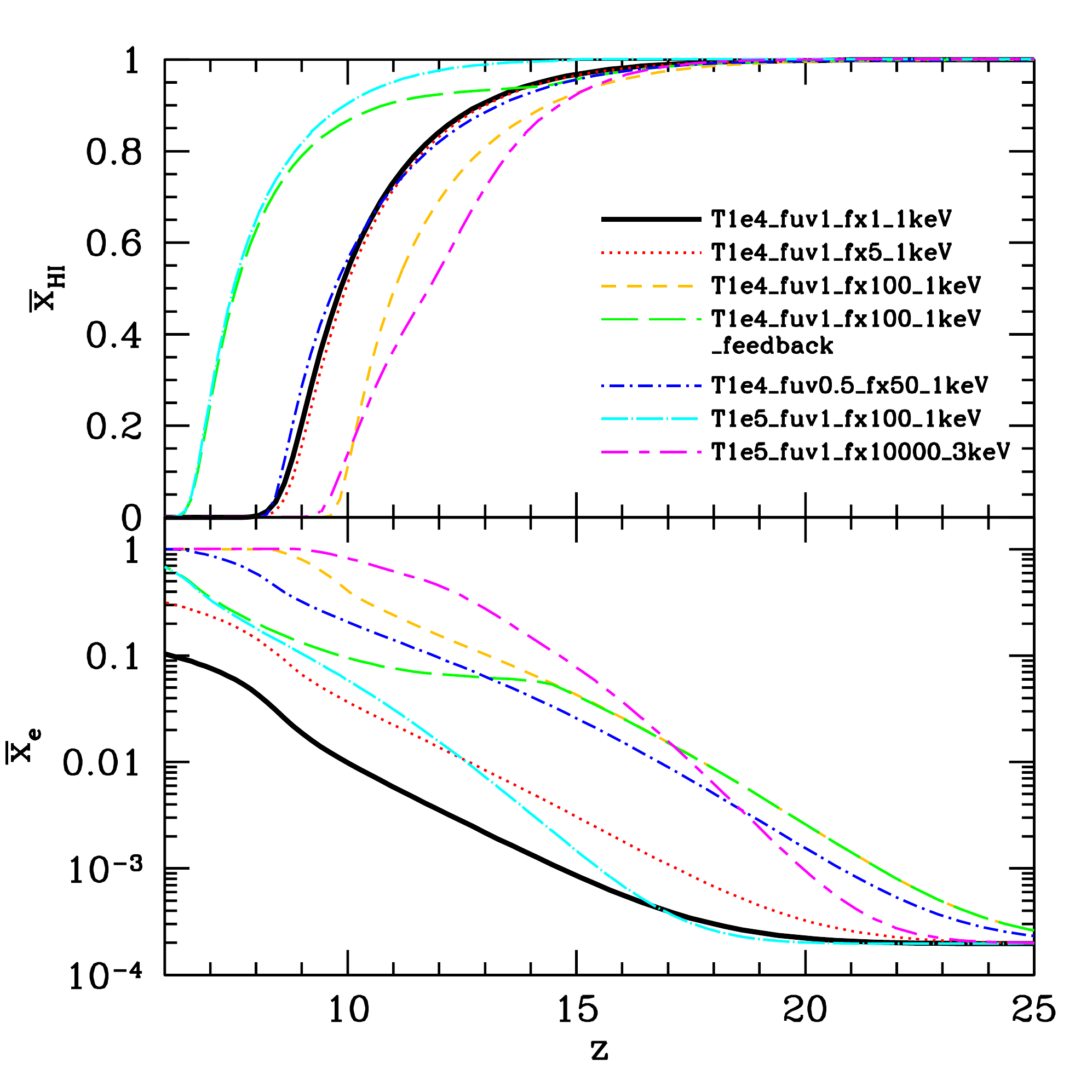}
}
\caption{
{\it Top panel:} Reionization histories of the runs listed in Table \ref{tbl:runs}. {\it Bottom panel:} The mean ionization fraction, $\bar{x}_e$, generated only by X-ray ionizations (ignoring the UV contribution except to compute the IGM opacity to X-rays).
\label{fig:reion_hist}
}
\vspace{-1\baselineskip}
\end{figure}

In the top panel of Fig. \ref{fig:reion_hist} we plot the reionization histories of our runs.  Again, there is little change from the fiducial model ({\it black solid curve}) as the X-ray efficiency is increased by a factor of $\approx5$ ({\it dotted red curve}). However, increasing the X-ray efficiency by a factor of 100 ({\it orange short-dashed curve}) causes reionization to happen earlier, with the midpoint shifting by $\Delta z \approx 1$.

Increasing the contribution of X-rays to reionization also causes reionization to happen more gradually, as can be seen when comparing the blue dot--short dashed and black solid curves, which are normalized to have the same $\tau_e$ (c.f. the top and fifth panels in Fig. \ref{fig:xH_lightcone}).  This effect is most seen in the extreme case of \highxray, where reionization is dominated by X-rays.  In this case, $d\avenf/dz$ is roughly constant after $\avenf\lsim 0.8$, and the reionization history doesn't show the ``knee'' feature where the slope increases in the middle stages as the collapse fraction grows.  This is due to the fact that as X-ray driven reionization progresses, the increasing abundance of galaxies is countered by the decreasing efficiency of X-ray ionization:  $f_{\rm ion, j} \rightarrow 0$ as $\avenf \rightarrow 0$, so only primary ionizations from X-ray photons can complete reionization (e.g. \citealt{SvS85, FS09, VEF11}).

Aside from directly ionizing the IGM, the heat input from early X-rays can result in thermal feedback: raising the effective Jeans mass required for gas to efficiently accrete onto DM halos (see the discussion in \S \ref{sec:runs}).  If this process is important, the first generations of galaxies would sterilize themselves, and reionization would be delayed until more massive DM halos emerge (see the fourth panel in Fig. \ref{fig:xH_lightcone}). The extreme limit of this scenario is explored in \xrayfeed\ ({\it green long dashed curve}), and results in a reionization history with a tail of $\avenf\approx90$\% extending to high-$z$, as also noted by \citet{RO04}.

In the bottom panel of Fig. \ref{fig:reion_hist}, we show the mean ionization fraction generated only by X-ray ionizations (i.e. ignoring the contribution from UV photons).  In fiducial reionization scenarios driven by UV ionizations, $x_e$ corresponds to the electron fraction in the ``neutral'' IGM, outside of the almost fully-ionized HII regions local to the sources.
%  However, $x_e$ overestimates the X-ray contribution during the advanced stages of reionization as it is computed prior to the post-processing by UV photons (see \S \ref{sec:UV}).
%  Note that the filling factor of HII regions, $Q_{\rm HII}$, is still used to estimate the opacity of the IGM when computing the X-ray flux \citep{MFC11}.
  As $\avenf$ decreases, the resulting increase in the X-ray mean free path (see eq. \ref{eq:mfp}) is the likely cause of the slight steepening of $dx_e/dz$. Note that the total IGM ionized fraction (including the porosity of HII regions, $Q_{\rm HII}$), $\bar{x}_i \approx Q_{\rm HII} + (1-Q_{\rm HII}) x_e$, is used to estimate the opacity of the IGM when computing the X-ray flux \citealt{MFC11}.  Hence, the X-ray mean free path will increase rapidly during reionization as $Q_{\rm HII}$ increases.  \citet{Baek10} also noted that such a redshifting X-ray background can have a non-negligible contribution to X-ray driven reionization.  This effect is most evident in the models where $\Tvir\sim10^4$ K halos drive reionization;  on the other hand, more massive sources on the exponential tail of the mass function appear too suddenly to show this feature (e.g. \citealt{Lidz07}).

%%%%%%%%%%%%%%%%%%%%%%%%%%%%%%%%%%%%%%%%%%%%%%%%%%%%%%%%%%%%%%%%%%%%%
%%%%%%%%%%%%%%%%%%%%%%%%%%%%%%%%%%%%%%%%%%%%%%%%%%%%%%%%%%%%%%%%%%%%%

\subsection{Ionization Morphology}
\label{sec:morpho}

%%%%%%%%%%%%%%%%%%%%%%%%%%%%%%%%%%%%%%%%%%%%%%%%%%%%%%%%%%%%%%%%%%%%%
\begin{figure}
\vspace{-1\baselineskip}
{
\includegraphics[width=0.45\textwidth]{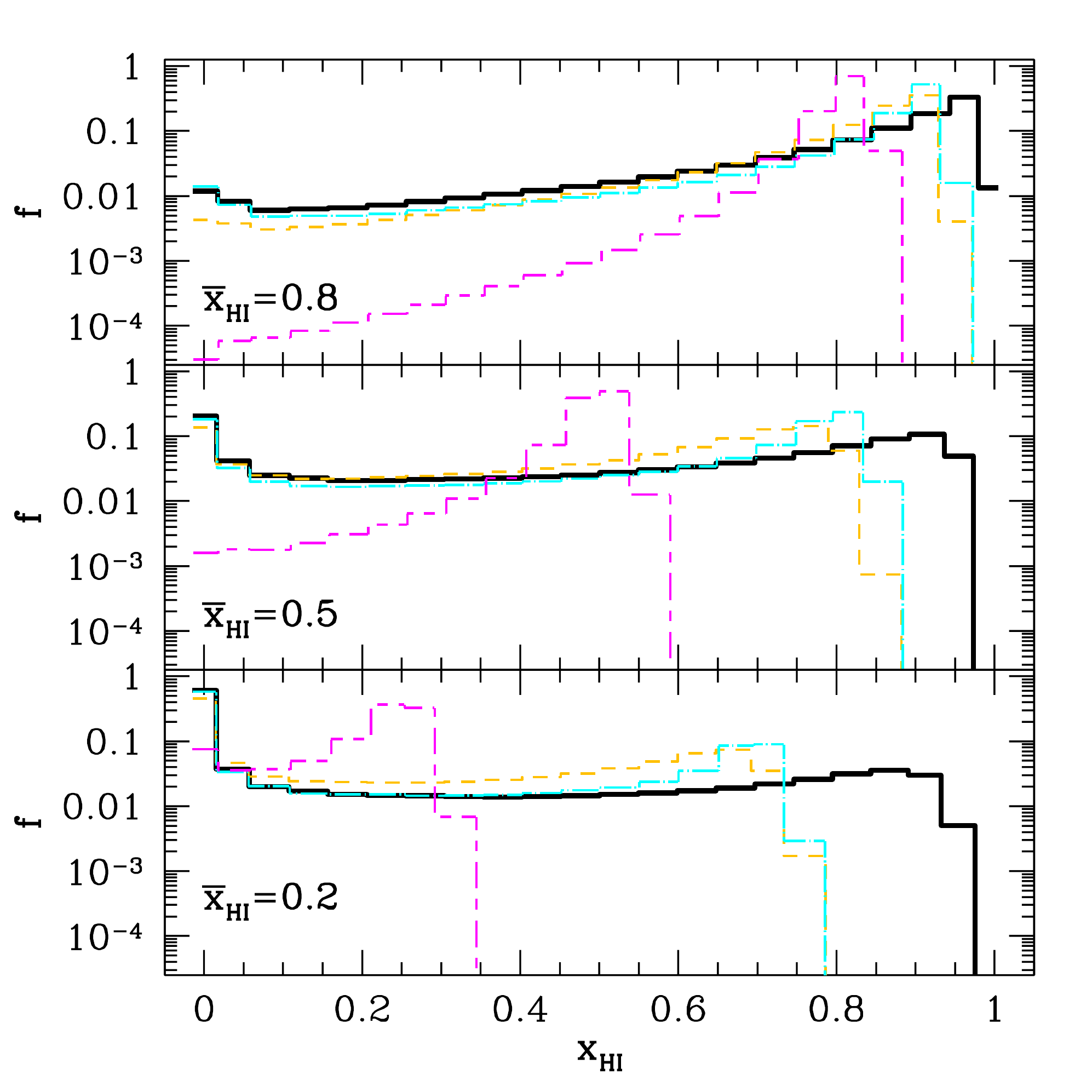}
}
\caption{
Fraction of $R_{\rm filter}= 3$ Mpc regions having a given $x_{\rm HI}$.  All smoothing is done with a top-hat filter. The runs are denoted by the same line styles as in Fig. \ref{fig:reion_hist}: \lowxray\ ({\it black solid}), \xray\ ({\it orange short-dashed}), \highTvir\ ({\it cyan dot long-dashed}),\highxray\ ({\it magenta short long dashed}). The panels correspond to $\avenf=0.8$, 0.5, 0.2 ({\it top to bottom}).
\label{fig:xH_PDFs}}
\vspace{-1\baselineskip}
\end{figure}
%%%%%%%%%%%%%%%%%%%%%%%%%%%%%%%%%%%%%%%%%%%%%%%%%%%%%%%%%%%%%%%%%%%%%

  It has been suggested that abundant X-rays in the early Universe could have a large impact on reionization morphology (e.g., see the discussion in \citealt{Haiman11}).  From Fig. \ref{fig:xH_lightcone}, we can see that this is the case only in extreme models, with no ``soft'' ($\lsim1$ keV) X-rays.  In this section we further quantify the impact of X-rays on reionization morphology.

  In Fig. \ref{fig:xH_PDFs}, we plot the histograms of the neutral fraction in $R_{\rm filter}= 3$ Mpc regions, at different stages of reionization.  The fiducial run ({\it black solid curve}) results in the most inhomogeneous ionization field since it has the smallest contribution of X-rays.

 As the X-ray efficiency is increased by a factor of 100 ({\it orange short-dashed curve}), the ionization field becomes noticeably more homogeneous. There is a significant suppression (a factor of $\approx$3) in the number of fully ionized regions in the early stages ($\avenf\approx0.8$).  At a fixed global neutral fraction, $\avenf$, including X-rays decreases the relative contribution of fully-ionized HII regions (driven mostly by UV sources).  Furthermore, the histogram becomes noticeably more peaked with a sharp reduction in the number of highly-neutral regions.  Specifically, there are no 3 Mpc regions having $\nf \gsim 0.95, 0.85, 0.75$ at $\avenf\approx 0.8,0.5, 0.2$, respectively.

As sources are hosted by rarer, more biased halos, the reionization field becomes slightly more inhomogeneous (compare the orange and cyan curves).  However, this difference is not as large as the one imprinted by X-ray reionization, since the DM halo bias increases only by several tens of percent going from $\Tvir=10^4$ K to $\Tvir=10^5$ K \citep{McQuinn07}.  {\it This result indicates that a significant contribution ($\sim$tens of percent) of X-rays to reionization results in a more homogeneous reionization morphology which cannot be countered by having more biased sources.}

The magenta curve in Fig. \ref{fig:xH_PDFs} corresponds to the ``extreme'' model, \highxray, in which X-rays complete reionization. The sources in this model also have an SED with a more energetic mean photon energy of $h\bar{\nu}\approx3$ keV.  As could be guessed from Fig. \ref{fig:xH_lightcone}, the reionization morphology is this model is significantly more homogeneous, with the ionization histograms sharply peaked around $\avenf$.

%%%%%%%%%%%%%%%%%%%%%%%%%%%%%%%%%%%%%%%%%%%%%%%%%%%%%%%%%%%%%%%%%%%%%
\begin{figure}
\vspace{-1\baselineskip}
{
\includegraphics[width=0.45\textwidth]{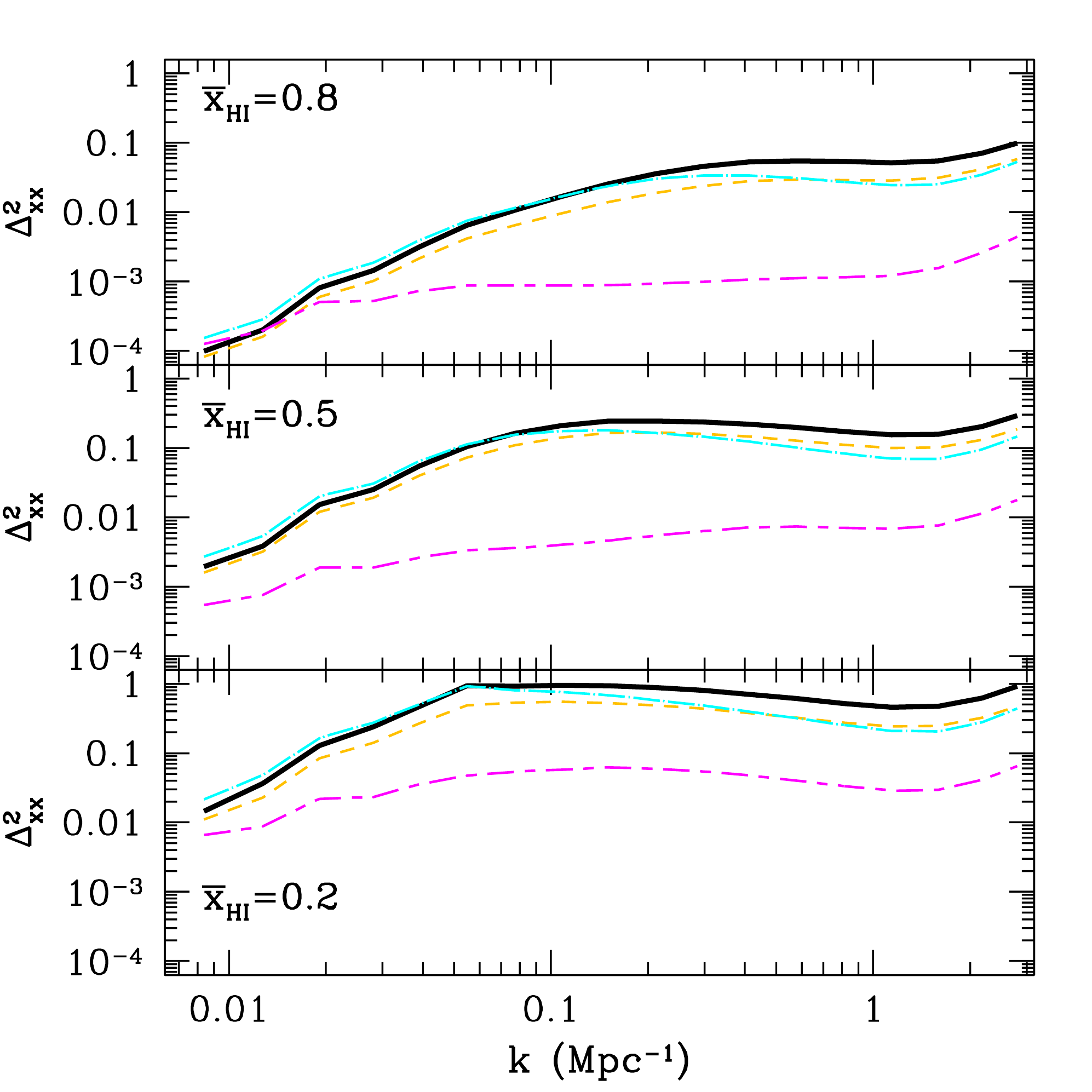}
}
\caption{
Power spectra of the ionization fields for the same models and epochs shown in Fig. \ref{fig:xH_PDFs}.
\label{fig:xH_ps}
}
\vspace{-1\baselineskip}
\end{figure}
%%%%%%%%%%%%%%%%%%%%%%%%%%%%%%%%%%%%%%%%%%%%%%%%%%%%%%%%%%%%%%%%%%%%%

In Fig. \ref{fig:xH_ps} we plot the ionization field power spectra ($\Delta^2_{\rm xx} = k^3/(2\pi^2 V) ~ \langle|\delta_{\rm xx}|^2\rangle_k$, with $\delta_{\rm xx}=x_{\rm HI}/\avenf - 1$) for the same models and epochs shown in Fig. \ref{fig:xH_PDFs}.  The same trends discussed above can also be seen in this figure.  The fiducial model displays the standard ``knee'' feature, which shifts to larger scales imprinting the HII region size in the early stages of reionization, and either the matter power spectrum or the mean free path of UV photons (set by Lyman limit systems) in the late stages of reionization (e.g. \citealt{AA12}).  This ``knee'' feature should be one of the fundamental observables of the upcoming 21cm interferometry measurements.% (e.g. \citealt{Lidz08}).

Increasing the X-ray contribution to reionization to $\sim$50\% (comparing the orange and black curves) results in a roughly scale-free suppression of the power spectrum by a factor of $\lsim2$.  However the ``knee'' feature is still preserved.  Further moving the sources to more biased halos (comparing the orange and blue curves) results in a more peaked bubble distribution, with a suppression of small-scale power and a boost in large-scale power.  We note that {\it the small-scale power is most sensitive to the X-rays, while the large scale power is most sensitive to the halos which host the sources} (e.g. \citealt{McQuinn07}).  On the other hand, the extreme model with X-ray driven reionization results in a very large (1--2 orders of magnitude) suppression of power, especially on small-scales.

%%%%%%%%%%%%%%%%%%%%%%%%%%%%%%%%%%%%%%%%%%%%%%%%%%%%%%%%%%%%%%%%%%%%%
\begin{figure*}
\vspace{+0\baselineskip}
{
\includegraphics[width=0.45\textwidth]{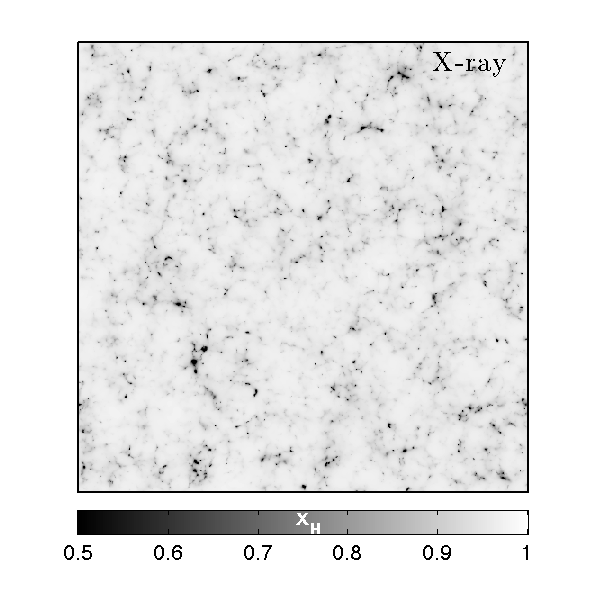}
\includegraphics[width=0.45\textwidth]{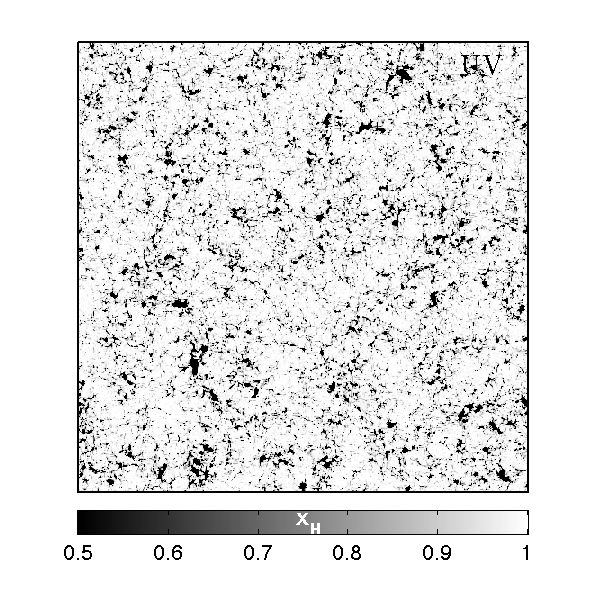}
\includegraphics[width=0.45\textwidth]{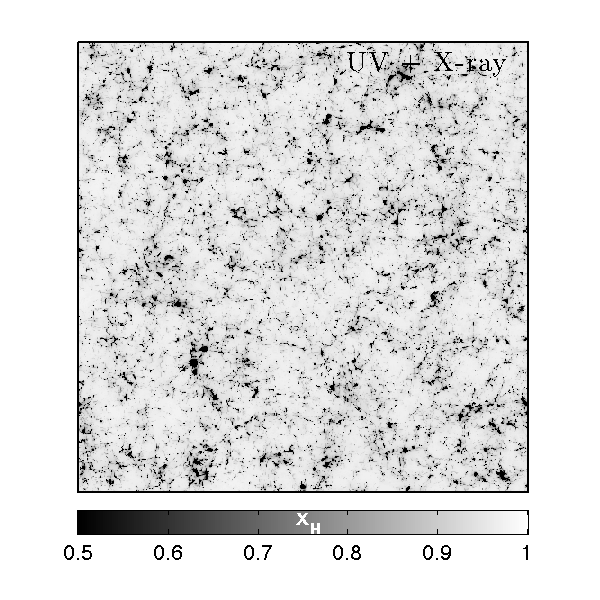}
\includegraphics[width=0.45\textwidth]{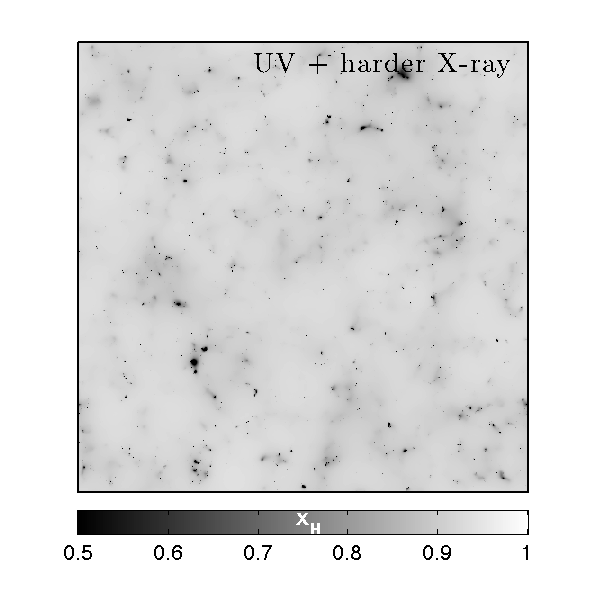}
}
\caption{
Slices through ionization fields at $\avenf=0.9$. The first panel shows just X-ray ionizations in the \matchtau\ model (i.e. 1-$x_e$ computed with eq. \ref{eq:ion_rateacc}). The second panel corresponds to ionizations from UV photons in the same model, but without any X-ray sources.  The third panel is the total ionization field in the \matchtau\ model.  The last panel is the ionization field in the \highxray\ run, where soft X-rays below $\lsim0.9$ keV are obscured. All slices are 1.5 Mpc deep, and show only the 0.5$<\avenf<$1 range for improved contrast.
\label{fig:xH_slices}
}
\vspace{-1\baselineskip}
\end{figure*}

These trends are qualitatively evident in Fig. \ref{fig:xH_slices}, where we show slices through the ionization fields at $\avenf=0.9$.  The first panel shows just ionizations from X-rays in the \matchtau\ model (turning off the UV sources).  The second panel shows the just the ionizations resulting from UV photons in the same model (turning off the X-ray sources).  It is evident that X-rays result in a smoother ionization field.  However, even without any UV sources, the soft X-rays are able to carve out highly ionized regions surrounding the first, most highly biased sources.

The third panel shows the total ionization field in the same model, resulting from both X-ray and UV sources (note that this field is not just the sum of the first two, but instead corresponds to an earlier epoch since all panels are chosen at $\avenf\approx0.9$).  We see that small ionization structures are most affected by X-rays, as also noted in Fig. \ref{fig:xH_ps}.  When compared at the same $\avenf$, the ionizations contributing to the small, late appearing, UV-driven HII regions, effectively shift to the uniform, partially ionized, X-ray driven ``haze''.  Larger HII regions are less affected, since these correspond to highly-biased regions hosting the first X-ray sources.  In these regions the X-rays themselves pre-ionized a large fraction of the HI, making it easier for the UV photons to finish the job. 

 The fourth panel corresponds to the \highxray\ model.  Unlike the model shown in the first panel, the \highxray\ model contains no soft X-rays with energies below $0.9$ keV.  The higher energy photons in this extreme model understandably create a more uniform ionization field.

%%%%%%%%%%%%%%%%%%%%%%%%%%%%%%%%%%%%%%%%%%%%%%%%%%%%%%%%%%%%%%%%%%%%%%%%%%%%%%%%%%%%%%%%%%%%%%%%%%%%%%%%%%%%%%%%%%%%%%%%

\subsection{The kinetic Sunyaev-Zel'dovich signal from inhomogeneous reionization}
\label{sec:kSZ}

One of the most promising near-term probes of reionization is the kSZ signal.  The kSZ results from the scattering of cosmic microwave background (CMB) photons off of inhomogeneities in the ionization and velocity fields  \citep{SZ80, OV86, Vishniac87, MF02}.  The kSZ can be decomposed into the post-reionization, so-called Ostriker-Vishniac (OV; \citealt{OV86, MF02}) component, and a component from during reionization referred to as the patchy kSZ \citep{GH98, KSD98}.   Observational efforts with the Atacama Cosmology Telescope (ACT) and the South Pole Telescope (SPT) are poised to detect this signal for the first time, with projected 1 $\muKK$-level sensitivity to the dimensionless kSZ power spectrum around a multipole of $l=3000$, $\Ptot$.  $l=3000$ is the ``sweet-spot'', corresponding to scales small enough not be dominated by the primary anisotropies and large enough not to be dominated by the CIB.  As an integral measurement, $\Ptot$ is sensitive to the duration of the epoch during which ionized structures were of comparable size ($l=3000$ corresponds to $\approx 20$ Mpc at high-$z$).

Indeed, recent SPT measurements place a bound of $\Ptot<2.8~\muKK$ at 95\% confidence limit (C.L.), which degrades to $\Ptot<6~\muKK$ if a significant correlation between the tSZ and the CIB is allowed \citep{Reichardt12}.  \citet{MMS12} showed that the $\Ptot<2.8~\muKK$ constraint is {\it inconsistent with all physically-motivated, UV-driven reionization scenarios, even with conservatively low estimates of the OV contribution.}  Reionization by UV photons is simply too inhomogeneous on $\approx 20$ Mpc scales to accommodate this constraint, with the patchy kSZ signal contributing at least 1.5 $\muKK$.

This implies that either: (i) there is a significant correlation between the CIB and the tSZ; and/or (ii) the early stages of reionization occurred in a much more homogeneous manner, perhaps driven by X-ray photons with large mean free paths.  Upcoming combined analyses of microwave and Herschel far infrared data will likely test the former scenario.  Here we explore the latter.

The patchy kSZ signal can be decreased by: (i) lowering the redshift of reionization; (ii) shortening its duration; and/or (iii) decreasing ionization structure on $l=3000$ scales\footnote{Note that a measurement of the patchy kSZ power combined with $\tau_e$ does {\it not} uniquely determine the redshift and duration of reionization, since the ionization morphology affects the shape of the kSZ power, in addition to its amplitude \citep{MMS12}.}.  X-ray driven reionization does the latter.

\begin{figure}
\includegraphics[width=0.45\textwidth]{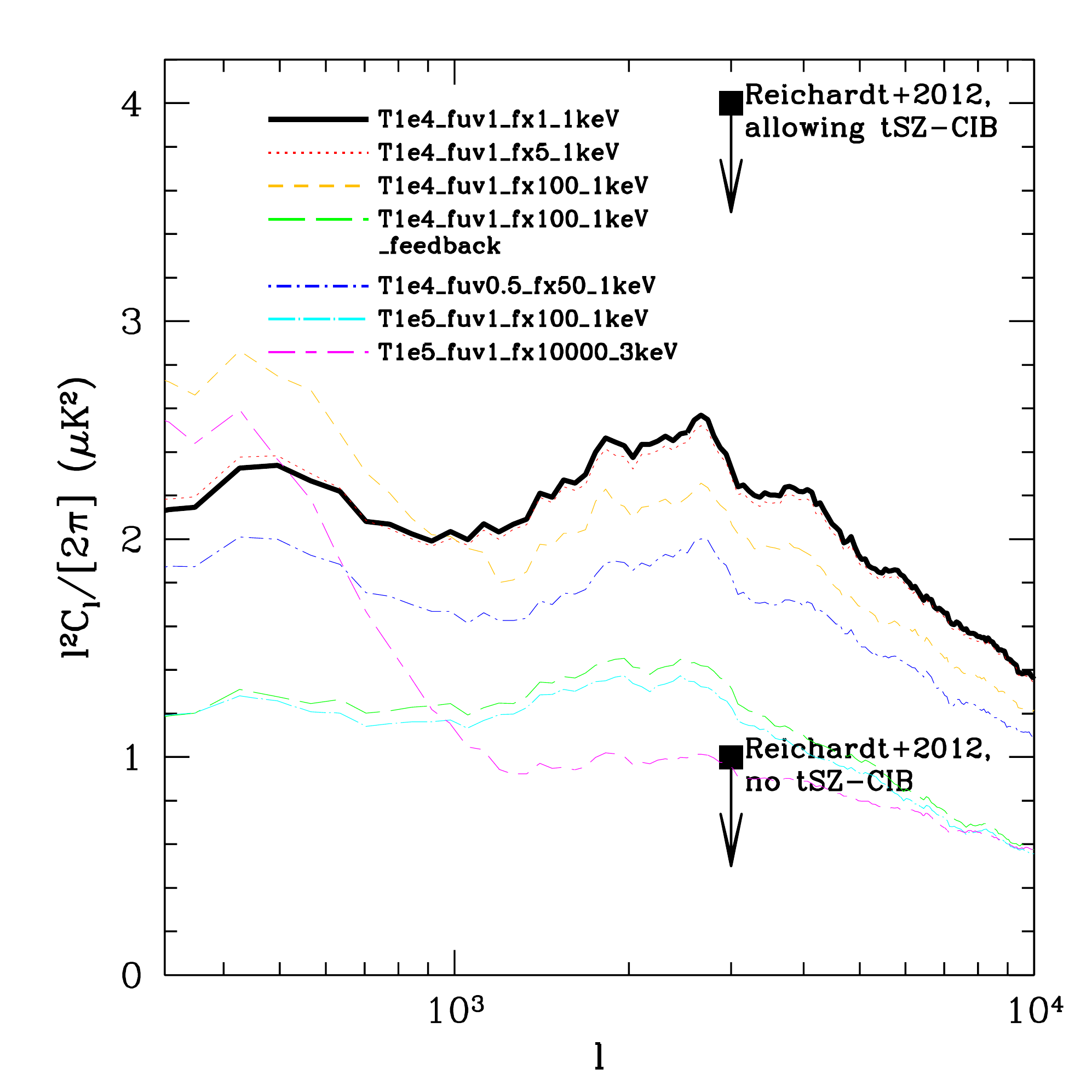}
\caption{
\label{fig:kSZ}
The kSZ power spectrum sourced from $z>5.6$.  The points denote the recent 95\% C.L. on the $l=3000$ power by \citet{Reichardt12}, assuming no tSZ-CIB correlation ({\it bottom point}) and allowing for a tSZ-CIB correlation ({\it upper point}).  The points also include a conservatively-low contribution of $\approx2\muKK$ from the $z<5.6$ OV signal (e.g. \citealt{SRN12, MMS12}).
}
\vspace{-1\baselineskip}
\end{figure}

In Fig. \ref{fig:kSZ} we plot the kSZ angular power spectrum $C_{l}^{\rm kSZ} \equiv T_{\rm cmb}^2 |\tilde{\delta_T}(k)|^2$, where $T_{\rm cmb} = 2.73~$K and $\tilde{\delta_T}$ is the Fourier transform of
\begin{equation}
\label{eq:kSZ}
\delta_T \equiv \frac{\Delta T}{T}(\uunit) = \sigma_{\rm T} \int dz \,c\, (dt/dz) \, e^{-\tau_e(z)} n_e {\bf \uunit \cdot v}  ~ ,
\end{equation}
\noindent and $\uunit$ is the LOS unit vector, $\sigma_{\rm T}$ is the Thomson scattering cross section, $\tau_e(z)$ is the Thomson optical depth to redshift $z$ in the direction $\uunit$, ${\bf v}(\uunit, z)$ is the peculiar velocity, and $n_e(\uunit, z)$ is the electron number density.  Following \citet{MMS12}, we only show the power sourced from $z>5.6$, corresponding roughly to the lower limit on the end of reionization \citep{MMF11}.  Power spectrum values at $l=3000$ from our models are also shown in Table 1.  The fiducial model has $l=3000$ kSZ power from $z>5.6$ of $2.3\muKK$.

Interestingly, we see that X-rays have only a modest impact on the shape of the power spectrum at $l\gsim1000$, unless they dominate reionization (magenta curve).  This is in line with what we saw in the previous section: the effects on the morphology at fixed $\avenf$ are rather small.

As before, increasing the X-ray efficiency by a factor of $\approx5$ from the fiducial value has little effect on this observable.  When the X-ray efficiency is further increased to account for $\approx1/2$ of ionizations, and the reionization history is matched to the same $\tau_e$, the $l=3000$ patchy kSZ power is decreased by $\approx0.5\muKK$ (compare the solid black and blue dot dashed curves).
  On the other hand, if the UV and X-ray efficiencies are not decreased to match the same $\tau_e$ as the fiducial model, the same relative X-ray contribution ($\sim$ 1/2 of ionizations) results in only a $\approx0.2\muKK$ decrease in patchy kSZ power over the fiducial model (compare the solid black and orange short dashed curves), on scales smaller than $l\approx1000$ ($\sim$60 Mpc)\footnote{Behavior on larger scales is difficult to interpret due to cosmic variance in the integrand of eq. (\ref{eq:kSZ}): $n_e {\bf \uunit \cdot v}$ (\citealt{Jelic10}; Jeli\'c et al., in preparation).  However, these scales are dominated by the primary anisotropies, and so are impossible to be used for the kSZ in the near future.}.   In this case, the more uniform component resulting from the X-rays is counteracted by shifting reionization to earlier epochs.  An earlier reionization increases the kSZ signal in Fig. \ref{fig:kSZ} through: (i) an increase in the mean density of the IGM in the integrand of eq. \ref{eq:kSZ}; and (ii) a larger post-reionization (OV) component to the signal down to $z=5.6$, the redshift cut we use.

\begin{figure}
\includegraphics[width=0.45\textwidth]{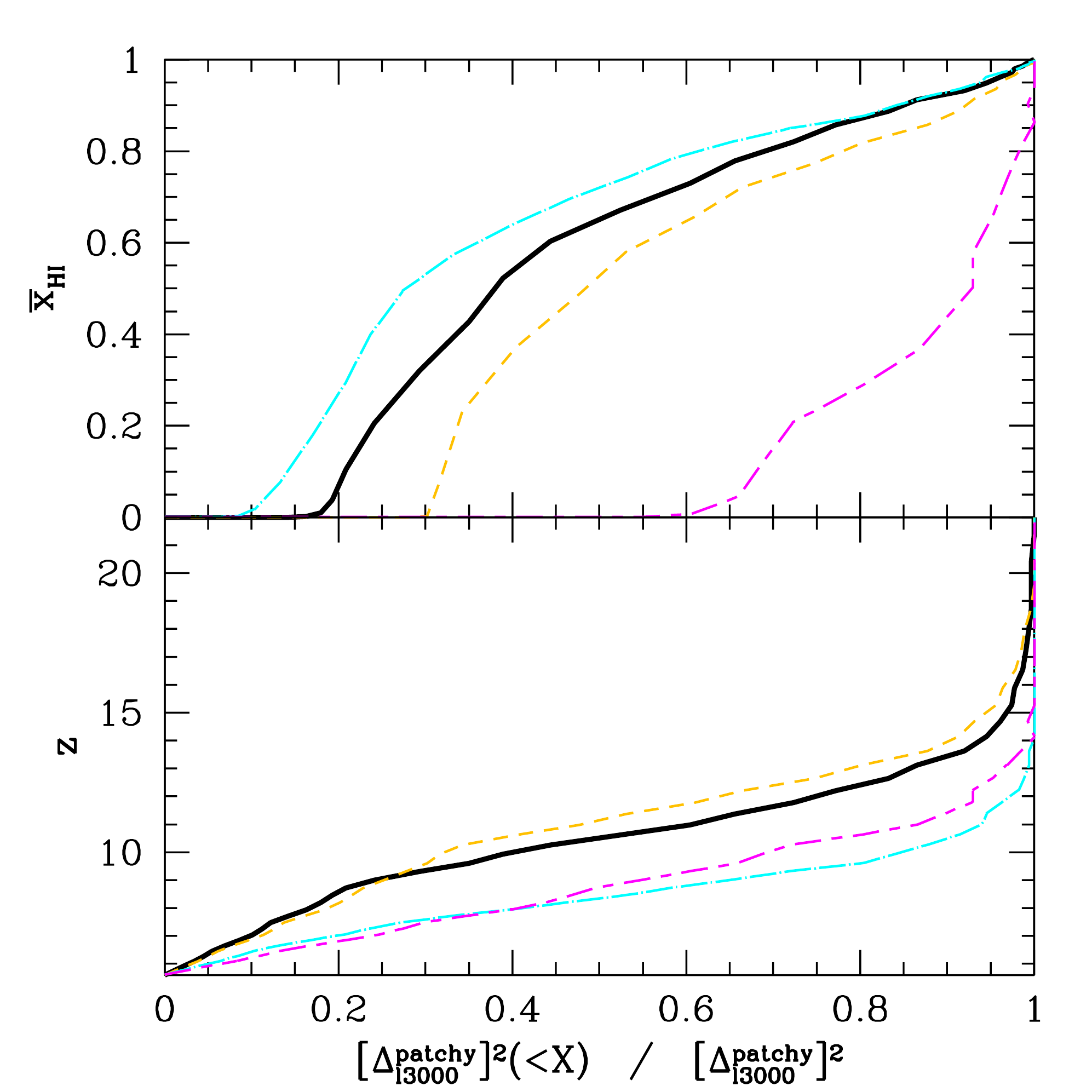}
\caption{
\label{fig:kSZ_dist}
The cumulative fraction of the $z>5.6$ kSZ signal at $l=3000$ sourced by redshifts less than $z$ ({\it bottom panel}) and corresponding global neutral fractions less than $\avenf$ ({\it top panel}).  We show the same models as in Fig. \ref{fig:xH_PDFs}.
}
\vspace{-1\baselineskip}
\end{figure}

This is illustrated in the top panel of Fig. \ref{fig:kSZ_dist}, where we show the cumulative fraction of the $z>5.6$ kSZ signal at $l=3000$ sourced at global neutral fractions less than $\avenf$.  In the fiducial model, $\approx$20\% of the total $z>5.6$ power comes from post-reionization (OV) $5.6<z\lsim8$, and the other 80\% comes from the epoch of reionization (patchy kSZ).  When the X-ray efficiency is increased by a factor of 100, these fractions change to 30\% (70\%) for the OV (patchy) kSZ components. However, the shape of this cumulative density distribution (CDF) is essentially unchanged.  On the other hand, as we increase $\Tvir$, a larger fraction of the power is sourced from the early stages of reionization.  As noted in \citet{MMS12}, increasing $\Tvir$ results in a more uniform HII bubble size, reducing the small-scale structure.  In the later stages of reionization, it is the small-scale structure (the tail of the HII bubble distribution), which dominates the power on $l=3000$ ($\approx$ 20Mpc) scales.  This is further evidence that using $\Ptot$ to place limits on the end of reionization is highly model-dependent.

%%%%%%%%%%%%%%%%%%%%%%%%%%%%%%%%%%%%%%%%%%%%%%%%%%%%%%%%%%%%%%%%%%%%%
%%%%%%%%%%%%%%%%%%%%%%%%%%%%%%%%%%%%%%%%%%%%%%%%%%%%%%%%%%%%%%%%%%%%%

%%%%%%%%%%%%%%%%%%%%%%%%%%%%%%%%%%%%%%%%%%%%%%%%%%%%%%%%%%%%%%%%%%%%%
%%%%%%%%%%%%%%%%%%%%%%%%%%%%%%%%%%%%%%%%%%%%%%%%%%%%%%%%%%%%%%%%%%%%%

Understandably, \highxray\ has the largest impact on the patchy kSZ signal.  In fact, it is the only one of our models which is marginally consistent with the recent aggressive SPT bounds on $\Ptot$.  From the top panel of Fig. \ref{fig:kSZ_dist}, we see that 60\% of the $z>5.6$ power in the model just comes from the OV effect post-reionization, $5.6<z\lsim9$.  Since this model is already unrealistically extreme (see the discussion in \S \ref{sec:runs}), we conclude that X-rays could not by themselves match the recent aggressive bound from SPT.  {\it There must be a sizable contribution from the tSZ-CIB cross-correlation}.

%%%%%%%%%%%%%%%%%%%%%%%%%%%%%%%%%%%%%%%%%%%%%%%%%%%%%%%%%%%%%%%%%%%%%%%%%%%%%%%%%%%%%%%%%%%%%%%%%%%%%%%%%%%%%%%%%%%%%%%%%
%%%%%%%%%%%%%%%%%%%%%%%%%%%%%%%%%%%%%%%%%%%%%%%%%%%%%%%%%%%%%%%%%%%%%
\begin{figure*}
\vspace{+0\baselineskip}
{
\includegraphics[width=0.9\textwidth]{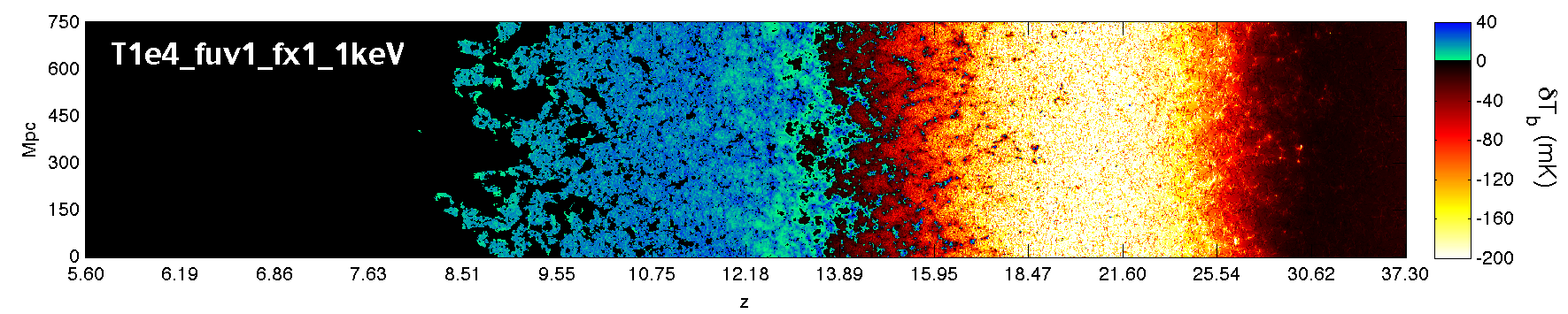}
\includegraphics[width=0.9\textwidth]{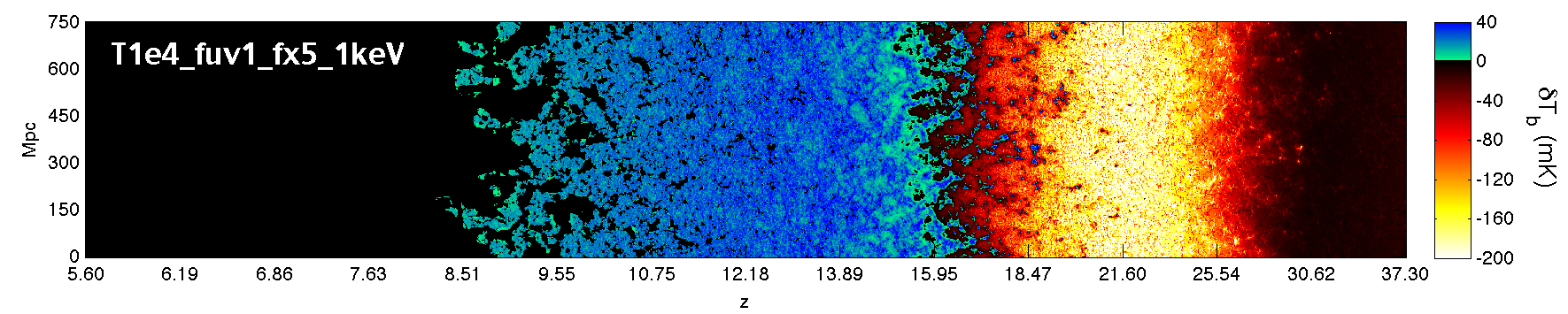}
\includegraphics[width=0.9\textwidth]{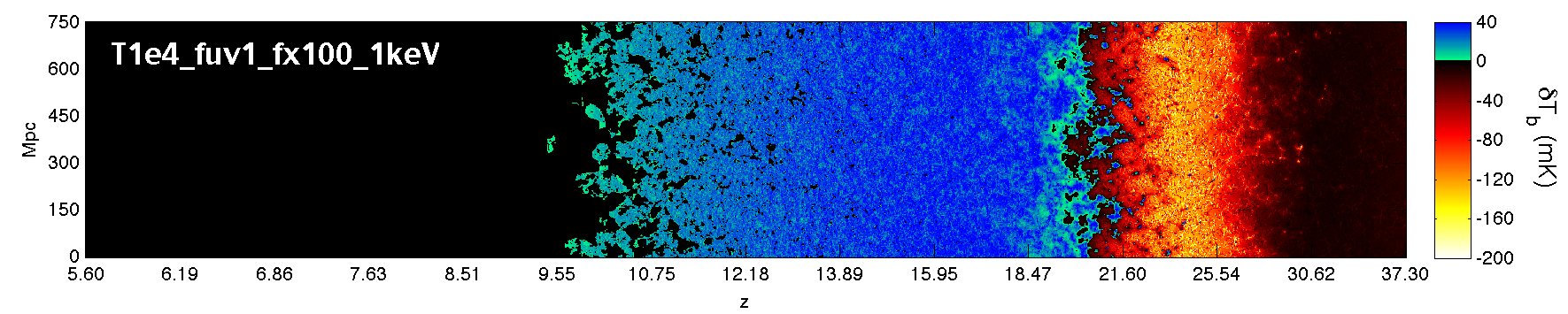}
\includegraphics[width=0.9\textwidth]{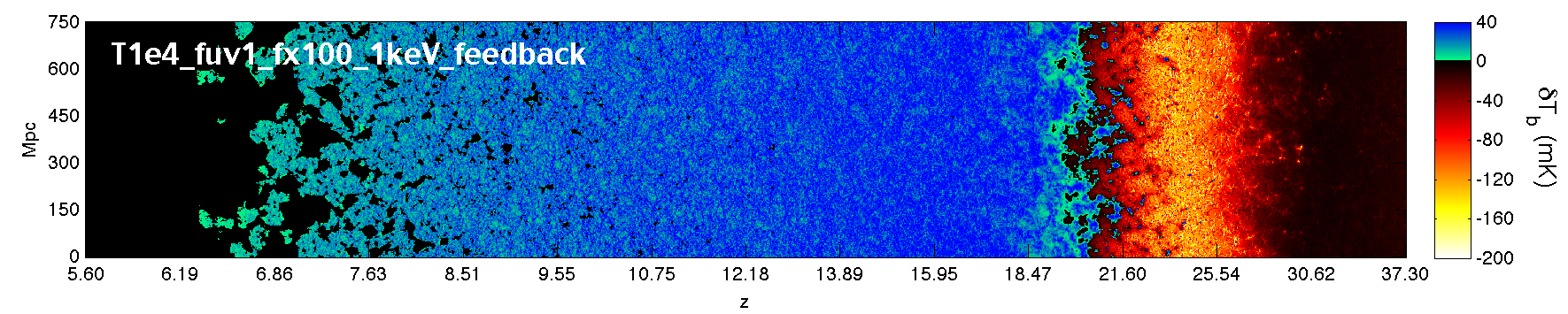}
\includegraphics[width=0.9\textwidth]{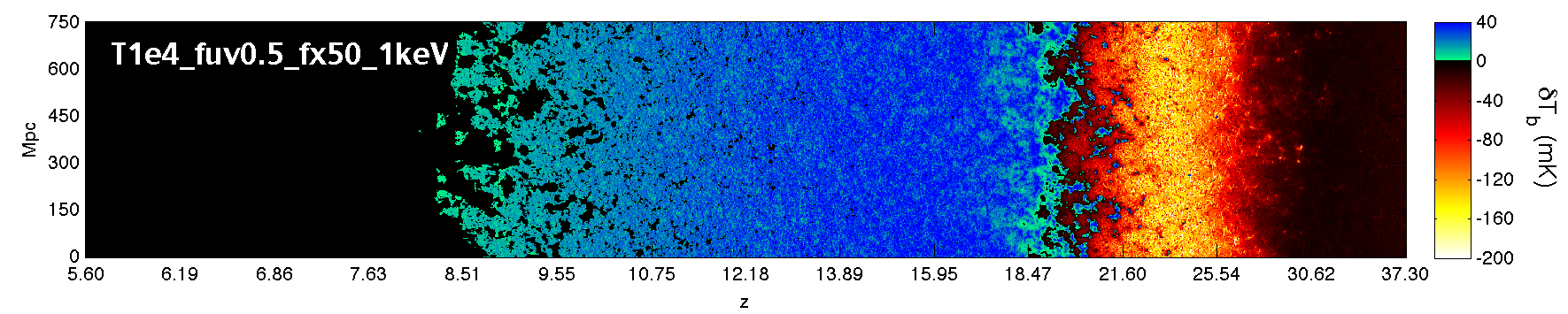}
\includegraphics[width=0.9\textwidth]{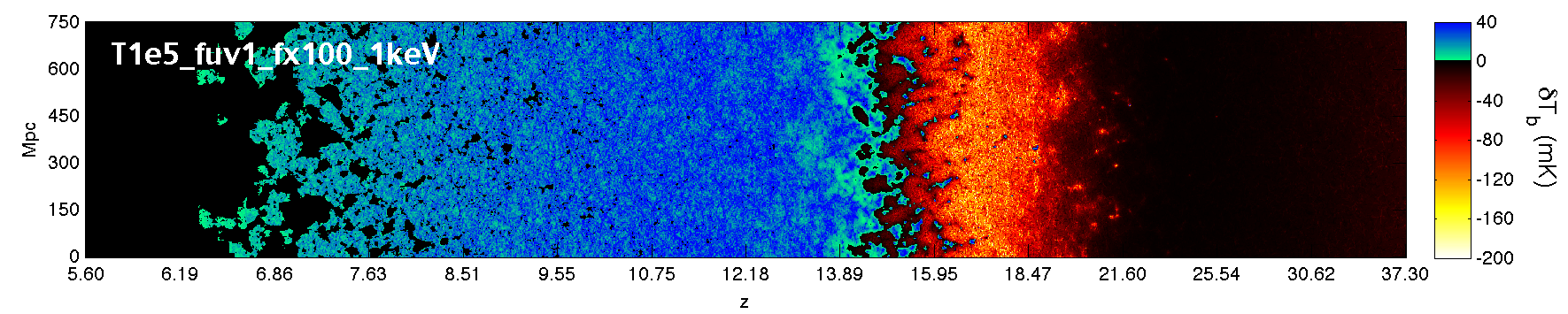}
\includegraphics[width=0.9\textwidth]{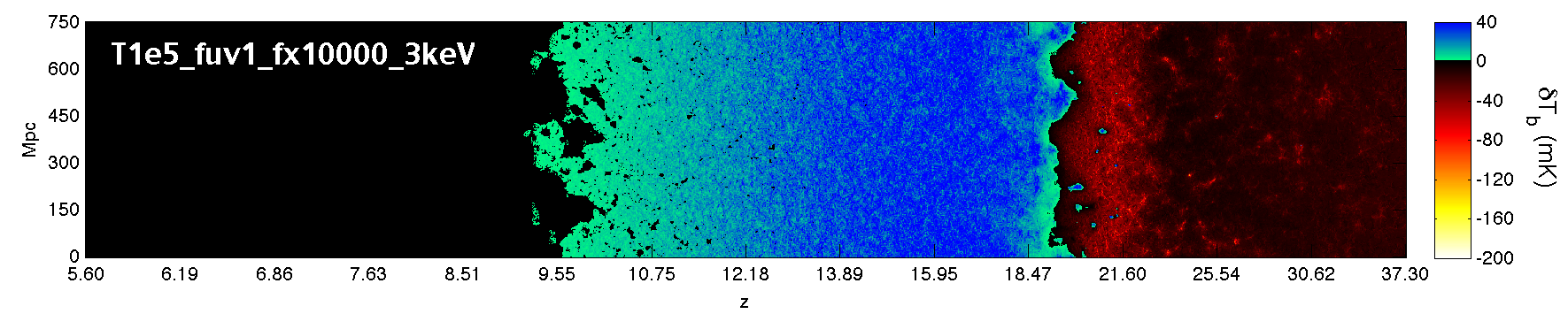}
}
\caption{
Slices through the predicted 21cm signal for our models. The slices show evolution with comoving distance along the light cone (x-axis).  For movies of some of these models, please see http://homepage.sns.it/mesinger/21cm\_Movie.html.
\label{fig:21cm_slices}
}
\vspace{-1\baselineskip}
\end{figure*}
%%%%%%%%%%%%%%%%%%%%%%%%%%%%%%%%%%%%%%%%%%%%%%%%%%%%%%%%%%%%%%%%%%%%%

\subsection{Cosmological 21cm Signal}
\label{sec:21cm}

The most promising probe of early X-rays is probably the redshifted 21cm line from IGM neutral hydrogen.  
 Precursor instruments like EDGES have already ruled out extremely rapid ($\Delta z\lsim0.06$) reionization models from the all-sky signal \citep{BR10}.
 First generation interferometers, like the Low Frequency Array (LOFAR; \citealt{Harker10})\footnote{http://www.lofar.org/},  Murchison Wide Field Array (MWA; \citealt{Tingay12})\footnote{http://www.mwatelescope.org/}, and the Precision Array for Probing the Epoch of Re-ionization (PAPER; \citealt{Parsons10})\footnote{http://astro.berkeley.edu/$\sim$dbacker/eor/} are coming on-line, promising to measure the 21cm power spectrum during $7\lsim z \lsim 11$.  Second generation instruments, such as the Square Kilometer Array (SKA; \citealt{SKA12})\footnote{http://www.skatelescope.org/} and the Lunar University Network for Astrophysical Research (LUNAR)\footnote{http://lunar.colorado.edu}, should have the sensitivity to directly image the topology of reionization (low signal-to-noise, large-scale images might even be possible with first-generation interferometers; \citealt{Zaroubi12}).  In addition to reionization, {\it the 21cm signal is also a sensitive probe of the thermal history of the IGM pre-reionization, when early X-rays have the largest imprint} (e.g. \citealt{Furlanetto06}).

The 21cm signal is usually represented in terms of the offset of the 21cm brightness temperature from the CMB temperature, $\Tcmb$, along a line of sight (LOS) at observed frequency $\nu$ (c.f. \citealt{FOB06}):

\begin{align}
\label{eq:delT}
\nonumber \delT(\nu) = &\frac{\Ts - \Tcmb}{1+z} (1 - e^{-\tau_{\nu_0}}) \approx \\
\nonumber &27 \nf (1+\delNL) \left(\frac{H}{dv_r/dr + H}\right) \left(1 - \frac{\Tcmb}{\Ts} \right) \\
&\times \left( \frac{1+z}{10} \frac{0.15}{\Omega_{\rm M} h^2}\right)^{1/2} \left( \frac{\Omega_b h^2}{0.023} \right) {\rm mK},
\end{align}
 
\noindent where $T_S$ is the gas spin temperature, $\tau_{\nu_0}$ is the optical depth at the 21-cm frequency $\nu_0$, $\delNL({\bf x}, z) \equiv \rho/\bar{\rho} - 1$ is the evolved (Eulerian) density contrast, $H(z)$ is the Hubble parameter, $dv_r/dr$ is the comoving gradient of the line of sight component of the comoving velocity, and all quantities are evaluated at redshift $z=\nu_0/\nu - 1$.

The spin temperature interpolates between the CMB temperature and the gas kinetic temperature, $T_K$.  Since the observation uses the CMB as a backlight, a signal is only obtained if $T_S \rightarrow T_K$.  This coupling is achieved through either: (i) collisions, which are effective in the IGM at high redshifts, $z\gsim50$; or (ii) a Lyman alpha background [so-called Wouthuysen-Field (WF) coupling; \citealt{Wouthuysen52, Field58}], effective soon after the first sources turn on.  The later coupling mechanism is the relevant one for the epochs in this work.

The Lyman alpha background has two main contributors:  (1)  X-ray excitation of HI, which scales with the X-ray intensity, eq. (\ref{eq:J}), but is computed with the corresponding fraction of photon energy going into \lya, $f_{\rm Ly\alpha}$ (taken from \citealt{FS09}); and (2) stellar emission of photons in the Lyman bands, which is computed assuming Population II stellar emission spectra and summing over the Lyman resonance backgrounds \citep{BL05_WF}.  We assume that the stellar emission is sourced from the same halos as the X-ray emission (i.e. using the same $\Tvir$), and take $f_\ast=0.1$.  For more details on these calculations, please see \citep{MFC11}.

In Fig. \ref{fig:21cm_slices}, we show slices through the $\delT$ fields in our models.  It is immediately obvious that the 21cm signal is a physics-rich probe, encoding information on various processes during and before reionization.  Although the exact timing varies, all of our models show the same, fiducial sequence (c.f. \citealt{Furlanetto06}; \S 2.1 in \citealt{MO12}):
\begin{packed_enum}
\item {\bf Collisional coupling}: The IGM is dense at high redshifts, so the spin temperature is uniformly collisionally coupled to the gas kinetic temperature, $T_K = T_S \lsim \Tcmb$. Following thermal decoupling from the CMB ($z\lsim300$), the IGM cools adiabatically as $T_K \propto (1+z)^2$, faster than the CMB, $\Tcmb \propto (1+z)$.  Thus $\delT$ is negative.  This epoch, serving as a clean probe of the mater power spectrum at $z\gsim100$, is not shown in Fig. \ref{fig:21cm_slices}.
\item {\bf Collisional decoupling}:  The IGM becomes less dense as the Universe expands.  The spin temperature starts to decouple from the kinetic temperature, and begins to approach the CMB temperature again, $T_K < T_S \lsim \Tcmb$.  Thus $\delT$ starts rising towards zero.  Decoupling from $\Tk$ occurs as a function of the local gas density, with underdense regions decoupling first. Eventually (by $z\sim25$), all of the IGM is decoupled and there is little or no signal.  This epoch corresponds to the black region in the far right of the panels in Fig. \ref{fig:21cm_slices}.
\item {\bf WF coupling (i.e. Ly$\alpha$ pumping)}:  The first astrophysical sources turn on, and begin coupling $T_S$ and $T_K$, this time through the WF effect. $\delT$ becomes more negative, reaching values as low as $\delT\sim100$--200 (depending on the offset of the WF and X-ray heating epochs). This epoch corresponds to the black$\rightarrow$yellow transition\footnote{The Lyman $\alpha$ background has two main contributors: (i) X-ray excitation of HI; and (ii) direct stellar emission of photons between \lya\ and the Lyman limit.  In general, we find that either (ii) dominates, or the two are comparable.  However in our extreme model (see the bottom panel of Fig. \ref{fig:21cm_slices}), the direct excitation by X-rays is clearly driving the WF coupling, imprinting a very interesting spatial structure to the epoch.} in the panels of Fig. \ref{fig:21cm_slices}.
\item{\bf X-ray heating}: X-rays heat the IGM, with the spin temperature now coupled to the gas temperature, $T_K = T_S$. As the gas temperature surpasses $\Tcmb$, the 21cm signal changes from absorption to emission, becoming insensitive to the actual value of $T_S$ (see eq. \ref{eq:delT}).  This epoch corresponds to the yellow$\rightarrow$blue transition in the panels of Fig. \ref{fig:21cm_slices}.
\item {\bf Reionization}:  the IGM becomes ionized, a process which is inside-out on large scales.  The signal again approaches zero.  This epoch corresponds to the blue$\rightarrow$black transition in the panels of Fig. \ref{fig:21cm_slices}.
\end{packed_enum}

%%%%%%%%%%%%%%%%%%%%%%%%%%%%%%%%%%%%%%%%%%%%%%%%%%%%%%%%%%%%%%%%%%%%%
\begin{figure}
\vspace{-1\baselineskip}
{
\includegraphics[width=0.45\textwidth]{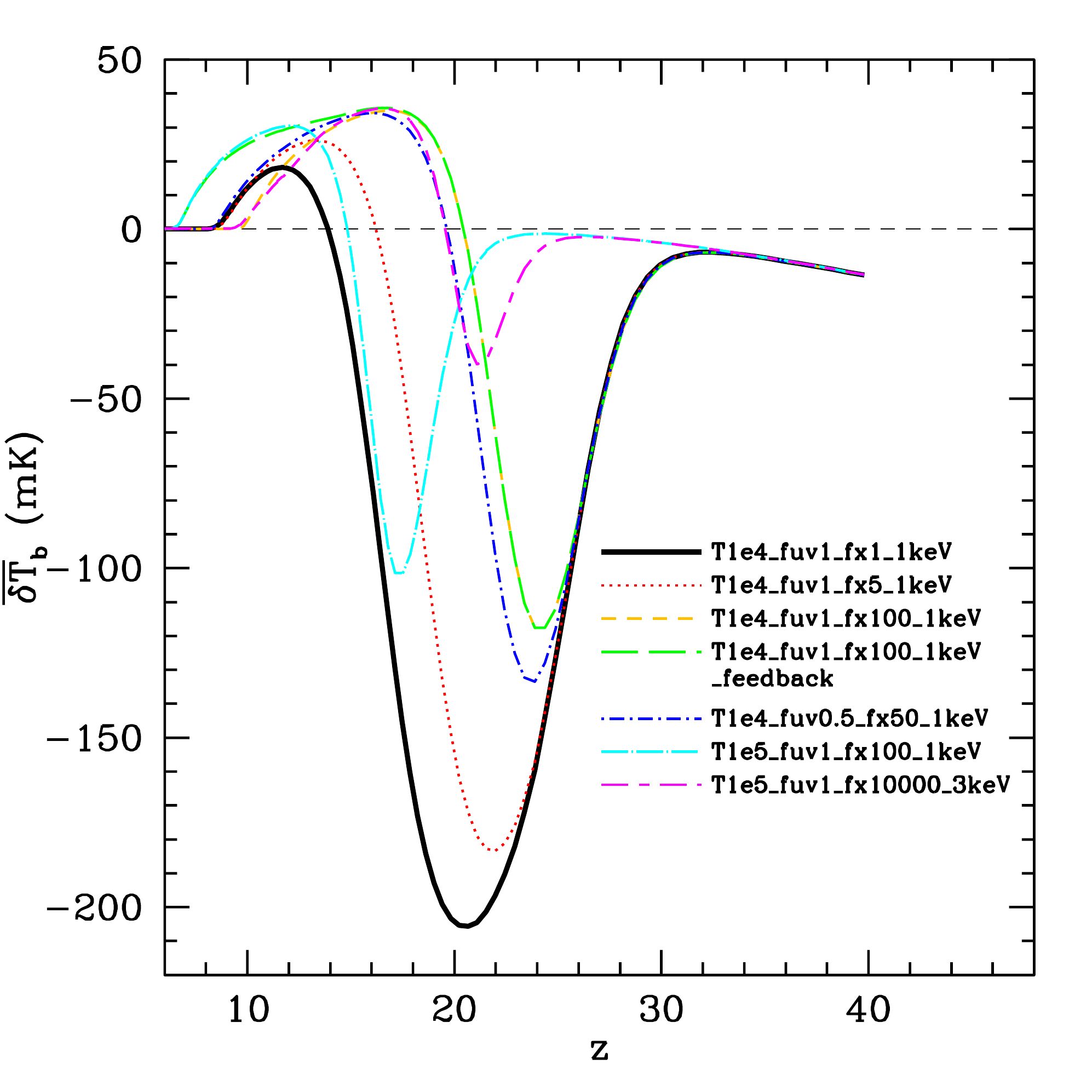}
}
\caption{
Redshift evolution of the average 21cm brightness temperature offset from the CMB.
\label{fig:delT_hist}
}
\vspace{-1\baselineskip}
\end{figure}
%%%%%%%%%%%%%%%%%%%%%%%%%%%%%%%%%%%%%%%%%%%%%%%%%%%%%%%%%%%%%%%%%%%%%

These stages can be seen in more quantitative detail in Fig. \ref{fig:delT_hist}, where we plot the corresponding evolution of the average $\delT$.  This figure clearly shows the emission peak and absorption trough.  The position and height/depth of the mean signal depends on the timing of the astrophysical milestones discussed above\footnote{Our fiducial model has a deeper $\bar{\delT}$ absorption trough and a lower emission peak than some previous estimates based on \citet{Furlanetto06}, due to our choice of a lower fiducial X-ray efficiency as discussed in \S \ref{sec:runs}.  A lower X-ray efficiency results in a wider WF coupling $\rightarrow$ X-ray heating transition, and a narrower X-ray heating $\rightarrow$ reionization transition.}.
 The photon production rates in our models trace the evolution of the collapse fraction, $f_{\rm coll}$.  Hence, if smaller dark matter (DM) halos host the first sources, then WF coupling, X-ray heating and reionization will occur earlier, and more gradually\footnote{Even more gradual heating can result in some DM annihilation models, since the heating is driven by the collapse fraction in $\gsim \Msun$ DM halos \citep{Valdes12}.}.

All other things being equal, the stronger the X-ray efficiency of the early sources, the shallower the $\delT$ absorption trough.  This is due to two reasons: (i) X-ray heating can begin before WF coupling is complete with $T_S > T_K$; (ii) the ratio $\Tcmb/T_K$ evolves roughly as $\propto (1+z)^{-1}$ prior to heating, so earlier X-ray heating results in a smaller $\delT$ contrast even if $T_S \approx T_K$. 

Following X-ray heating, the $(1-\Tcmb/T_S)$ term in eq. (\ref{eq:delT}) goes to unity and the IGM is seen in emission against the CMB.  If there is subsequently a delay before the onset of reionization, then only the cosmological terms in eq. (\ref{eq:delT}) are not unity, and the mean signal decreases as $\bar{\delT} \propto \sqrt{1+z}$.  Such an epoch would facilitate a clean measurement of the matter power spectrum at lower, more accessible redshifts than the Dark Ages.  We see in Fig. \ref{fig:delT_hist} that this would occur if reionization was driven by more massive sources than those driving X-ray heating, such as would be the case with strong thermal feedback (\xrayfeed; see also the fourth panel in Fig. \ref{fig:21cm_slices}).

%%%%%%%%%%%%%%%%%%%%%%%%%%%%%%%%%%%%%%%%%%%%%%%%%%%%%%%%%%%%%%%%%%%%%
\begin{figure}
\vspace{-1\baselineskip}
{
\includegraphics[width=0.45\textwidth]{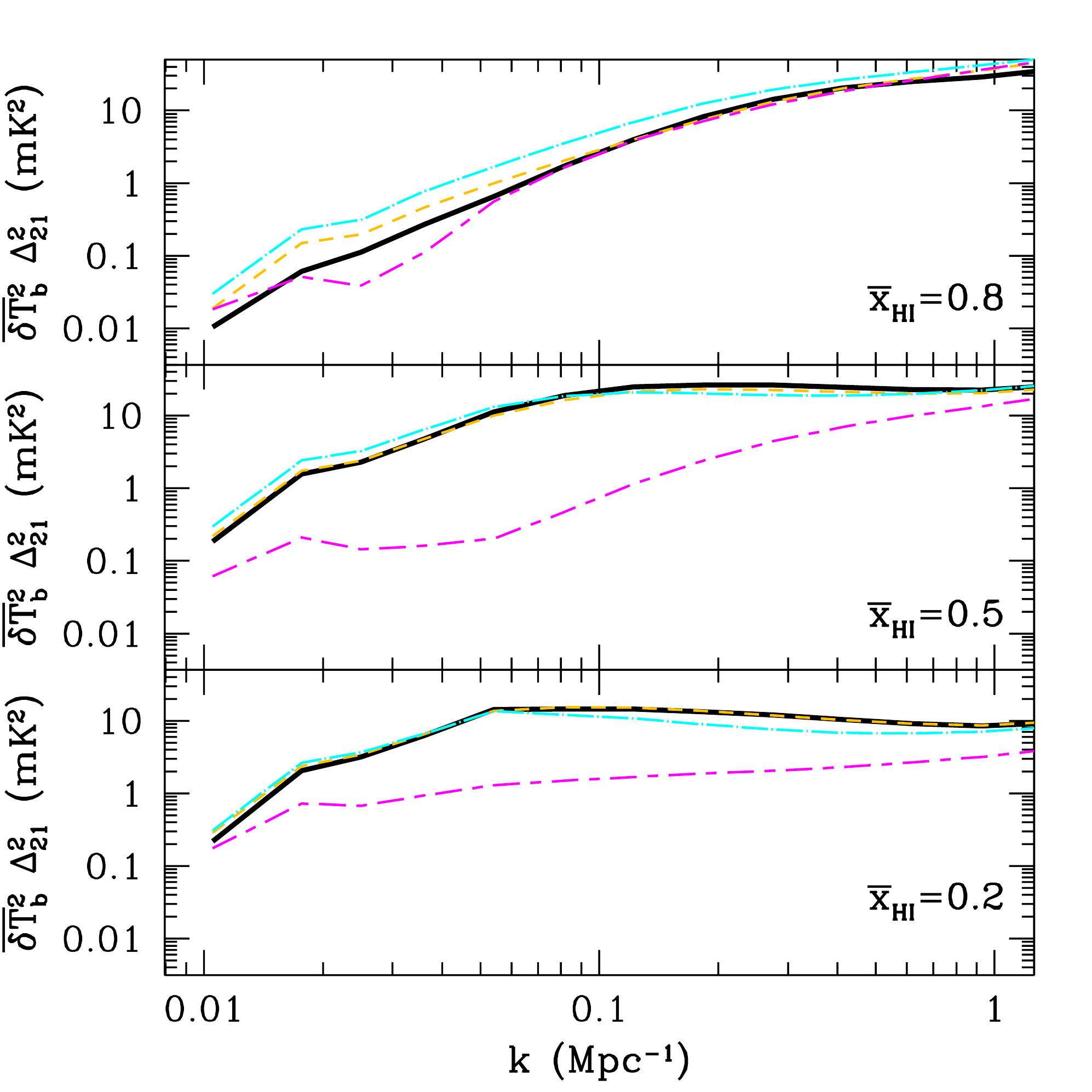}
}
\caption{
21cm power spectra from the same models as in Fig. \ref{fig:xH_PDFs}, shown at $\avenf=$ 0.8, 0.5, 0.2 ({\it top to bottom}).
\label{fig:ps21cm}
}
\vspace{-1\baselineskip}
\end{figure}
%%%%%%%%%%%%%%%%%%%%%%%%%%%%%%%%%%%%%%%%%%%%%%%%%%%%%%%%%%%%%%%%%%%%%

%%%%%%%%%%%%%%%%%%%%%%%%%%%%%%%%%%%%%%%%%%%%%%%%%%%%%%%%%%%%%%%%%%%%%
\begin{figure}
\vspace{-1\baselineskip}
{
\includegraphics[width=0.45\textwidth]{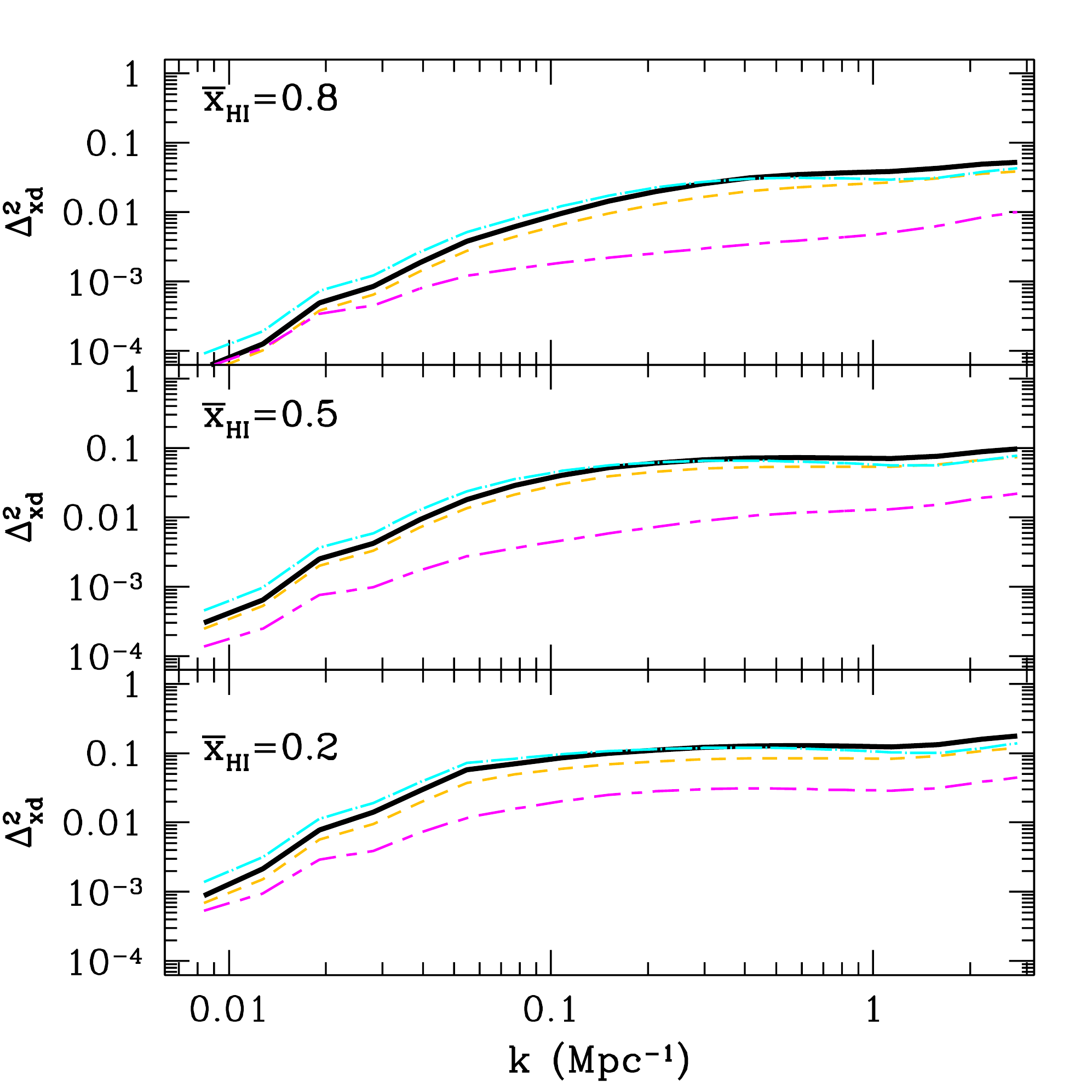}
}
\caption{
Ionization-density cross power spectrum for the same models as in Fig. \ref{fig:xH_PDFs}, shown at $\avenf=$ 0.8, 0.5, 0.2 ({\it top to bottom}).
\label{fig:XDps}
}
\vspace{-1\baselineskip}
\end{figure}
%%%%%%%%%%%%%%%%%%%%%%%%%%%%%%%%%%%%%%%%%%%%%%%%%%%%%%%%%%%%%%%%%%%%%

In Fig. \ref{fig:ps21cm}, we show the 21cm power spectra from the same models and epochs shown in Fig. \ref{fig:xH_ps}.  Some of the same qualitative trends as noted in Fig. \ref{fig:xH_ps} can also be seen in Fig. \ref{fig:ps21cm}; however the differences between the models are smaller.  This can be understood if we decompose the 21cm power spectrum (post-heating) to first-order:
\begin{equation}
\Delta^2_{21}=\bar{T}_b^2 \left[\Delta^2_{\rm xx} - 2\avenf \Delta^2_{\rm xd} + \avenf^2 \Delta^2_{\rm dd}\right],
\label{eq:pk21}
\end{equation}
where $\bar{T}_b$ is the average brightness temperature in regions with $\avenf = 1$, and $P$ is the power spectrum of the ionization (X) and density (D) fields.   As we saw in \S \ref{sec:morpho}, the more-uniform X-ray component decreases $\Delta^2_{\rm xx}$ on small scales.  However, this mild decrease is compensated by a decrease in the cross-correlation term, $\Delta^2_{\rm xd}$ (shown in Fig. \ref{fig:XDps}), which contributes negatively to the 21cm power.  The impact of the cross term is most evident in the early stages of reionization, where the fiducial model ({\it black curve}) has a factor of $\lsim 2$--3 less 21cm power at $k\lsim0.1$ Mpc$^{-1}$ than \xray\ ({\it orange curve}).

%%%%%%%%%%%%%%%%%%%%%%%%%%%%%%%%%%%%%%%%%%%%%%%%%%%%%%%%%%%%%%%%%%%%%
\begin{figure*}
\vspace{-1\baselineskip}
{
\includegraphics[width=0.45\textwidth]{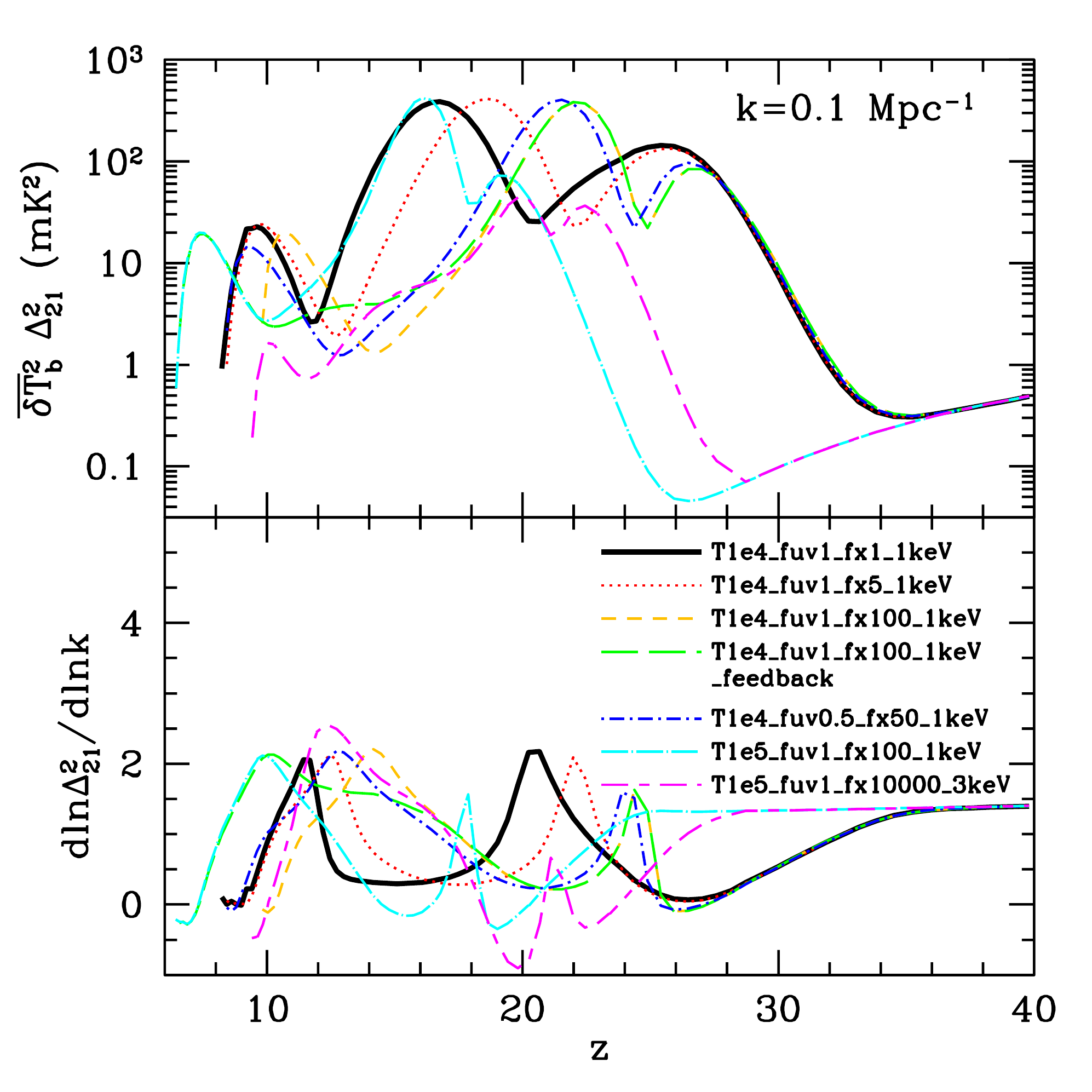}
\includegraphics[width=0.45\textwidth]{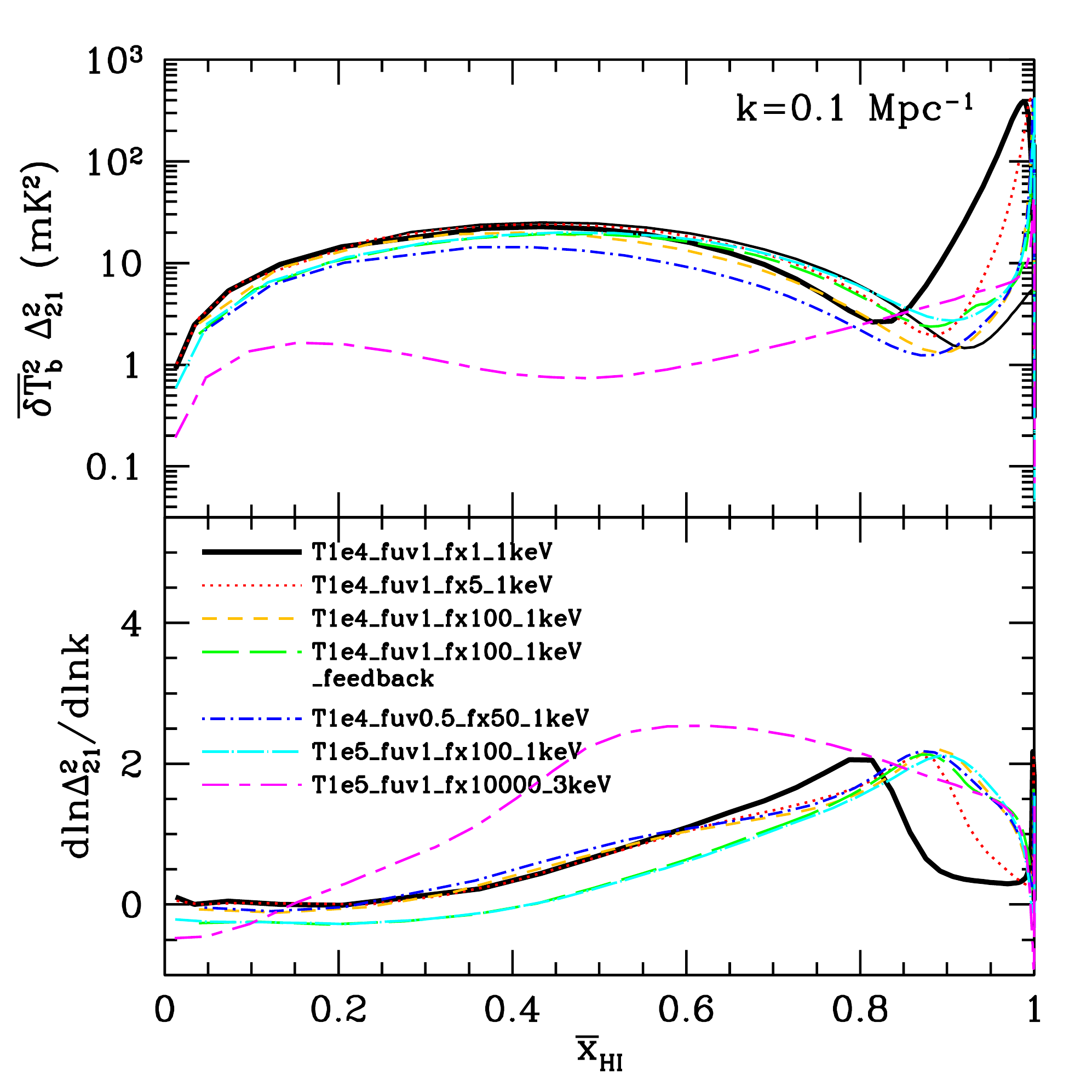}
}
\caption{
Evolution of the 21cm power ({\it top panels}) and its slope ({\it bottom panels}) at $k=0.1$ Mpc$^{-1}$.  Evolution is shown against redshift on the left, and against $\avenf$ on the right.  In the top right panel, we also show the power spectrum in the fiducial model, but computed with the commonly-used approximation of $T_S \gg \Tcmb$ ({\it thin, black solid curve}).
\label{fig:power_hist}
}
\vspace{-1\baselineskip}
\end{figure*}
%%%%%%%%%%%%%%%%%%%%%%%%%%%%%%%%%%%%%%%%%%%%%%%%%%%%%%%%%%%%%%%%%%%%%

In Fig. \ref{fig:power_hist}, we focus on the evolution of the $k=0.1$ Mpc$^{-1}$ mode, likely to lie in the middle of the decade of $k$-modes probed by the first generation interferometers (e.g. \citealt{Furlanetto09_21cmastro, Lidz08, Chapman12}).  The evolution of our fiducial model agrees very well with the numerical simulations of \citet{Baek10} (see the solid curves of the S6 model in their Fig. 6; this model has a similar X-ray efficiency as our fiducial model), but with their evolution shifted to later times (with their resolution they were unable to account for the large majority of atomically-cooled galaxies).

As discussed in previous works \citep{PF07, Baek10, MFC11}, the evolution of the $k=0.1$ Mpc$^{-1}$ large-scale power has three peaks, corresponding to the epochs of Ly$\alpha$ pumping, X-ray heating and reionization.  The earlier peaks (especially the X-ray heating one) are larger, due to the larger contrast available in the 21cm absorption regime.  The position and width of these peaks scale with the model parameters.  A higher $\Tvir$ results in delayed and more rapid transitions, tracing the growth of the exponential, high-mass tail of the halo mass function.  
Increasing the X-ray efficiency mainly shifts the X-ray heating peak to earlier epochs.  If X-ray heating occurs early enough (e.g. $f_{\rm X}\gsim10$), it overlaps with the Ly$\alpha$ pumping epoch, decreasing the associated peak in power.  Our extreme run, \highxray, generally has less power on $k=0.1$ Mpc$^{-1}$ scales throughout cosmic time, with the X-ray heating and WF coupling epochs overlapping significantly.  The hard X-rays driving both epochs create radiation fields with little small-scale structure, while fluctuations in both Ly$\alpha$ and gas temperature source the large-scale power; this results in a negative slope at $k=0.1$ Mpc$^{-1}$ during X-ray heating ({\it bottom panel}).

Zooming-in on the reionization epoch, in the right panel of Fig. \ref{fig:power_hist} we plot the same quantities as a function of $\avenf$.  In the top panel, we also show the power spectrum in the fiducial model, but computed with the commonly-used approximation of $T_S \gg \Tcmb$ ({\it thin, black solid curve}).
Our results are in excellent agreement with those in \citet{Lidz08}, who only considered reionization assuming $T_S \gg \Tcmb$  (see their Fig. 2).  \citet{Lidz08} note that the inside-out nature of reionization causes a drop in large-scale 21cm contrast at early times ($\avenf\gsim0.9$), before the growth of ionization structures pushes the power back up.  Stated differently in terms of eq. (\ref{eq:pk21}), the 21cm power before reionization (and assuming $T_S \gg \Tcmb$) traces the matter power spectrum, $\Delta^2_{\rm dd}$.  In the early stages when HII regions are small, the large-scale power drops due to the increase in the cross-correlation term, $\Delta^2_{\rm xd}$, which contributes negatively.  Then at $\avenf\lsim0.9$, the HII bubbles grow, dominating the 21cm power through the $\Delta^2_{\rm xx}$ term.  The $k=0.1$ Mpc$^{-1}$ 21cm power subsequently peaks at $\avenf\approx0.5$ \citep{Lidz08, Friedrich11}, with weak scale dependence (i.e. roughly zero slope; \citealt{Zahn11, MFC11}).  These trends are evident in the right panel of Fig. \ref{fig:power_hist}.

Including X-rays impacts this scenario in two main ways.  Firstly, the long mean free path of X-rays, combined with their lower ionization efficiency in regions with a higher ionized fraction, results in the uniform, partially-ionized ``haze'', seen most clearly in the bottom panel of Fig. \ref{fig:xH_lightcone}.  Therefore, models with a large fraction of X-ray ionizations have a weaker and slower growth of the $\Delta^2_{\rm xd}$ and $\Delta^2_{\rm xx}$ terms.  This results in a more extended initial stage of reionization (with the large-scale 21cm power falling as $\Delta^2_{\rm xd}$ increases, before it becomes dominated by the ionization structure, i.e. $\Delta^2_{\rm xx}$).  However, this effect is only really evident in the extreme \highxray\ model, where the $k=0.1$ Mpc$^{-1}$ power continues to fall until $\avenf\approx0.5$, roughly the epoch when $\Delta^2_{\rm xx} \sim \Delta^2_{\rm xd}$ (compare Figures \ref{fig:XDps} and \ref{fig:xH_ps}).   In the other models UV + softer X-ray photons drive the usual evolution of $\Delta^2_{21}$ described above, with the difference in power between the models being less than a factor of 2, even when X-rays contribute $\approx$1/2 of the ionizations.

Secondly, and more importantly, we see that {\it in realistic X-ray models the assumption of $T_S \gg \Tcmb$ is not valid in the early stages of reionization.}  Instead, the reionization and X-ray heating epochs overlap in virtually all of our models, resulting in a strong, scale-free drop in power at $\avenf\gsim0.8$--0.9.  This overlap is most notable in our fiducial model, which has the weakest X-ray heating.  Setting the $(1-\Tcmb/T_S)$ term to unity in eq. (\ref{eq:delT}) under-predicts the dimensional 21cm power spectrum in this model by one to two orders of magnitude at $\avenf \gsim 0.9$ (compare the thin and thick black solid curves in the top right panel).  Subsequently at $0.6 \lsim \avenf \lsim 0.8$, ignoring the spin temperature term over-predicts the signal (note that in our fiducial model: $\langle T_S \rangle \approx 10 \times \Tcmb$ at $\avenf\approx0.7$, decreasing $\bar{\delT}$ by $(1-\Tcmb/T_S)\approx90\%$).  The transition between over- and under- prediction using the common $T_S \gg \Tcmb$ assumption occurs somewhat later than the global transition from absorption to emission, $\langle T_S \rangle = \Tcmb$, as might be naively expected ($z\approx12$ rather than $z\approx14.7$).  This is due to the fact that, although $\bar{\delT} \sim 0$, the fluctuations in the gas temperature dominate the power spectrum around this epoch \citep{PF07}.
% Note that the peak of the mean brightness temperature, $\bar{\delT}$, in this model occurs at $z\approx12$, when reionization was already well underway,  $\avenf\approx0.8$.

\section{Conclusions}
\label{sec:conc}

Recent interest has focused on X-rays as a possibly significant contributor to cosmic reionization (e.g. \citealt{Haiman11, Mirabel11, FCV12}).  X-ray driven reionization would be fundamentally different from a UV driven one.  In particular, reionization could be more extended with a more uniform morphology.  In this work, we model the observational signatures of X-rays in the early Universe, attempting to provide an intuitive framework for interpreting upcoming observations.

We use the public code, \cmfast, to generate seven different models of reionization and pre-heating.  Our simulations are 750 Mpc on a side (currently unachievable by standard RT approaches hoping to capture the contribution from the dominant galaxy population), and span the redshift range $5.6<z<40$. We vary astrophysical parameters, looking for general trends and physical insight.  In addition to studying the timing and morphology of reionization, we also predict the kSZ and 21cm signatures in our models.

We find that by increasing the X-ray contribution, reionization occurs earlier, has less small-scale structure, and develops a partially-ionized homogeneous ``haze''.  However, when compared at the same $\avenf$, the impact is modest.  Namely, X-rays yield fewer small HII regions, with the IGM between HII regions being partially ionized.
%In general, however the impact is modest. When compared at the same $\avenf$, the ionizations contributing to the small, late appearing, UV-driven HII regions, effectively shift to the uniform, partially ionized, X-ray driven ``haze''.
  Larger HII regions are less affected, since these host the first, highly-biased X-ray sources.  In these regions the X-rays themselves pre-ionized a large fraction of the HI, making it easier for the UV photons to finish the job. Specifically, we find that models with a significant (tens of percent) contribution of X-rays to reionization exhibit: (i) a lack of fully neutral regions, and (ii) a suppression of small-scale ($k\gsim0.1$Mpc$^{-1}$) ionization power by a factor of $\lsim2$. These changes in the reionization morphology cannot be countered by having more biased sources (i.e. a higher $\Tvir$), and are therefore a robust indicator of an X-ray contribution to reionization.

The suppression of small-scale ionization structure by X-rays also results in a smaller kSZ signal at $l=3000$.  In particular, if X-rays contribute $\approx1/2$ of ionizations, the patchy kSZ power is decreased by $\approx0.5$ $\muKK$ with respect to a UV-only model having the same reionization history.  The shape of the $10^3 < l < 10^4$ power spectrum remains largely unaffected. The kSZ signal is reduced strongly enough to be marginally consistent with the recent aggressive SPT constraint \citep{Reichardt12}, only if X-rays fully reionize the Universe.  Since this model is already unrealistically extreme (see the discussion in \S \ref{sec:runs}), it is highly unlikely that X-ray reionization can by itself match this bound, which assumes no tSZ-CIB cross-correlation.  We therefore conclude that there must be a sizable contribution from the tSZ-CIB cross-correlation.

We note that the impact of X-rays on the kSZ signal could be distinguished from other astrophysical uncertainties only if UV reionization scenarios are unable to match the observed upper bound.  Since physically-motivated UV reionization models must contribute $\gsim 1.5 \muKK$ to the $l=3000$ kSZ power \citep{MMS12}, the tSZ-CIB correlation would have to contribute less than  $\lsim 0.5 \muKK$ in order for X-rays to be required to explain the observations (see Fig. \ref{fig:kSZ_dist}).

In general, X-rays have a small impact on the 21cm signal during the advanced stages of reionization ($\avenf\lsim0.7$), when compared at the same $\avenf$.  
The difference in 21cm power is less than a factor of $\approx$2, even when X-rays contribute $\approx$1/2 of the ionizations. The impact of X-rays on 21cm power is less than that on morphology, since the suppression of the ionization power is compensated by a decrease in the density-ionization cross-correlation (which contributes negatively to the 21cm power).  On the other hand, if harder X-rays complete reionization (our ``extreme'' model), the 21cm power could be lower by more than an order of magnitude in the later half of reionization, and there would be a local minimum in the evolution of large-scale power at $\avenf\approx0.5$ (see top right panel of Fig. \ref{fig:power_hist}).

The situation is different in the early stages of reionization as well as the pre-reionization epochs. 
During the pre-reionization epochs, X-rays govern the timing and duration of IGM heating.  The large-scale ($k\approx0.1$ Mpc$^{-1}$) 21cm power shows three maxima, corresponding to Ly$\alpha$ pumping, X-ray heating and reionization, with the X-ray heating one being the highest of the three.
During the early stages of reionization, X-rays can also have a large impact on the 21cm power spectrum.  This is primarily due to the overlap of the X-ray heating epoch and the early stages of reionization.  The overlap is strongest in our fiducial model, which has the weakest X-ray emissivity.  In particular, this overlap can lead to $\gsim10$--100 times more power at $k\approx0.1$Mpc$^{-1}$ than predicted with the commonly-used simplifying assumption of $T_S \gg \Tcmb$, at $\avenf\gsim0.9$. Subsequently, at $0.6 \lsim \avenf \lsim 0.8$, this assumption over-predicts the power by a factor of 1--2.  Therefore modeling X-ray heating is necessary to predict the 21cm signal even as late as the early stages of reionization ($\avenf \gsim 0.7$).

On the other hand, if thermal feedback was very efficient, the early X-ray emitting galaxies would self-sterilize by increasing the Jeans mass in the IGM, and reionization would be driven by more massive, later appearing sources (e.g. \citealt{RO04}). Such a scenario would result in an extended period after the completion of X-ray heating and before the beginning of reionization.  In this interim period, the 21cm line would be seen in emission, and would  offer a clean probe of the matter power spectrum.  This probe of cosmology occurs at much lower redshifts (making it easier to observe) than the commonly-implored Dark Ages at $z\gsim40$ (e.g. \citealt{Furlanetto09_21cmcosmo}).

In closing, we enumerate some quantitative observational signatures which can be used to constrain the first generations of X-ray sources:
\begin{packed_enum}
\item an X-ray contribution to reionization of $\gsim$ tens of percent would result in a lack of (almost) fully-neutral ($x_{\rm HI}\gsim0.9$) regions which are $\sim$ few cMpc in scale, during the last half of reionization;
\item a 21cm power spectrum tracing the matter power spectrum in emission can indicate strong thermal feedback from early X-rays;
\item if reionization is driven by $\sim$1 keV X-rays, we expect the large-scale ($k\approx0.1$Mpc$^{-1}$) 21cm power to fall until around $\avenf\approx0.5$, then rise again until $\avenf\approx0.2$, thereby creating a local minimum at $\avenf\approx0.5$ where standard models instead predict a local maximum (see the top panel of Fig. \ref{fig:power_hist}).
%a local minimum in the evolution of large-scale 21cm power around $\avenf\sim0.5$ is an indication of X-ray driven reionization;
\item if high-$z$ galaxies are more efficient at producing X-rays than local ones, the rise in 21cm power associated with X-ray heating would occur earlier, at $\avenf \gsim 0.9$.
%\item 21cm power at $k\approx0.1$ Mpc$^{-1}$ being lower at  $\avenf\approx0.95$  than the local maximum at $\avenf\approx0.5$, would indicate that the X-ray emissivity/SFR is stronger in the early Universe than at lower redshifts (by a factor of $\gsim5$).
\end{packed_enum}
Of the above, the most promising probe of X-rays in the early Universe would be 21cm measurements of the IGM heating epoch. The frequency coverage and sensitivity of the first generation interferometers is unlikely to yield a detection of X-ray heating, unless the X-ray efficiencies are smaller than our fiducial choice by a factor of $\gsim5$ (Ewall-Wice et al., in-preparation).  However, second generation interferometers like the SKA will be able to detect the epoch of X-ray heating, through the evolution of large-scale 21cm power (e.g. \citealt{Santos11, SKA12}). Of greater immediacy are upcoming, relatively inexpensive all-sky dipole experiments.  For example, the Large-Aperture Experiment to Detect the Dark Ages (LEDA\footnote{http://www.cfa.harvard.edu/LEDA}), scheduled to begin commissioning in 2013, will have a frequency coverage of 45-90 MHz ($z\approx$ 15--30).  These wide-beam experiments should detect the onset of X-ray heating through the global brightness temperature evolution (see Fig. \ref{fig:delT_hist}), provided that foregrounds are removed to the level of $\sim$tens of mK \citep{GH12}.

\vskip+0.5in

We thank Matthew McQuinn and Mark Dijkstra for comments on a draft version of this paper. DSS gratefully acknowledges support from NSF grant AST-0807444, the Keck Fellowship, and the Friends of the Institute.

\bibliographystyle{mn2e}
\bibliography{ms}

\end{document}